\documentclass[]{aa}  

\usepackage{graphicx}
\usepackage{txfonts}
\begin{document}

   \title{HCN-to-HNC intensity ratio:\\ a new chemical thermometer for the molecular ISM
   \thanks{Based on observations carried out with the IRAM30m Telescope. IRAM is supported by INSU/CNRS (France), MPG (Germany) and IGN (Spain).}} 


   \author{A. Hacar
          \inst{1}
          \and
          A. D. Bosman
          \inst{1}
          \and
          E. F. van Dishoeck
          \inst{1}
          }
          \institute{
         	   Leiden Observatory, Leiden University, P.O. Box 9513, 2300-RA Leiden, The Netherlands\\
         	   \email{hacar@strw.leidenuniv.nl}
             }

   \date{XXXX}

\abstract
{The gas kinetic temperature ($T_\mathrm{K}$) determines the physical and chemical evolution of the interstellar medium (ISM). However, obtaining reliable $T_\mathrm{K}$ estimates usually requires expensive observations including the combination of multi-line analysis and dedicated radiative transfer calculations.
}
{This work explores the use of HCN and HNC observations, and particularly the $I$(HCN)-to-$I$(HNC) intensity ratio ($I$(HCN)/$I$(HNC)) of their J=1--0 lines, as direct probe of the gas kinetic temperature in the molecular ISM.}
{We obtained a new set of large-scale observations of the HCN and HNC (1-0) lines throughout the Integral Shape Filament (ISF) in Orion. In combination with ancillary gas and dust temperature measurements, we find a systematic temperature dependence of the observed $I$(HCN)-to-$I$(HNC) intensity ratio throughout our maps. Additional comparisons with chemical models demonstrate that these observed $I$(HCN)/$I$(HNC) variations are driven by the effective destruction and isomerization mechanisms of HNC under low-energy barriers.}
{The observed variations of $I$(HCN)/$I$(HNC) with $T_\mathrm{K}$ can be described with  a two-part linear function. This empirical calibration is then used to create a temperature map of the entire ISF. Comparisons with similar dust temperature measurements in this cloud, as well as in other regions and galactic surveys, validate this simple technique for obtaining direct estimates of the gas kinetic temperature in a wide range of physical conditions and scales with an optimal working range between 15~K~$\lesssim  T_\mathrm{K} \le $~40~K. }
{Both observations and models demonstrate the strong sensitivity of the $I$(HCN)/$I$(HNC) ratio to the gas kinetic temperature. Since these lines are easily obtained in observations of local and extragalactic sources, our results highlight the potential use of this observable as new chemical thermometer for the ISM.} 


   \maketitle
%

\section{Introduction}\label{sec:intro}

The gas kinetic temperature ($T_\mathrm{K}$) represents the most fundamental thermodynamical property of the Interstellar Medium (ISM). In combination with the gas density $n$(H$_2$), the value of $T_\mathrm{K}$ determines the gas pressure ($P/k$=$n$(H$_2$)$T_\mathrm{K}$) and  thus the thermal support against gravity. $T_\mathrm{K}$ also sets the sound speed  ($c_s=\sqrt{(k~\mathrm{T}_K)/\mu}$) and the transition between the sonic and supersonic regimes. 
As a result, $T_\mathrm{K}$ sets the fragmentation scale of the star-forming gas in molecular clouds \citep{LAR85}. Moreover, the gas kinetic temperature regulates the chemical properties of the molecular gas that defines the activation and rate of gas-phase reactions \citep{VDI18}. Observationally, $T_\mathrm{K}$ also influences the excitation conditions and intensities of both atomic and molecular emission lines \citep[e.g.][]{SHI15}. Obtaining accurate and systematic measurements of $T_\mathrm{K}$ is therefore a key ingredient for describing the physical and chemical evolution of the ISM.

Different observational methods have been traditionally been employed to estimate the gas kinetic temperature in the molecular ISM using line and continuum observations.
$T_\mathrm{K}$ is regularly estimated from the analysis of the excitation temperatures (T$_{ex}$) and level populations \citep{GOL99} using multiple transitions of the same tracer such as CO ladders \citep[e.g.][]{PEN12}.
Similarly, $T_\mathrm{K}$ can be evaluated using temperature-sensitive transitions of either single tracers \citep[e.g. H$_2$CO;][]{MAN93,GIN16} or multiple isotopologues \citep[e.g. $^{12}$CO, $^{13}$CO, and C$^{18}$O; ][]{NIS15} in combination with radiative transfer models (e.g. large velocity gradients, LVG).
The gas kinetic temperature can also be inferred from the rotational temperatures ($T_\mathrm{rot}$) of collisionally excited transitions of density-selective tracers such as NH$_3$ \citep{HO83}. 
On the other hand, recent far-infrared (FIR) space observatories (i.e. {\it Herschel} and {\it Planck}) have popularized the use of dust temperature estimates ($T_\mathrm{dust}$) as a proxy of $T_\mathrm{K}$ using multiwavelength continuum observations \citep[e.g.][]{LOM14}.

The recent development of broad-band heterodyne receivers today enables routine production of spectral molecular maps with hundreds or thousands of individual beams in all types of ISM studies of protoplanetary disks \citep{JOR16}, molecular clouds \citep{PET16}, and nearby galaxies \citep{JIM19} using radiotelescopes such as the IRAM30m or the Atacama Large Millimeter Array (ALMA). 
While created for the study of individual sources, 
the use of the above temperature estimates in these large molecular datasets is nevertheless far from trivial. 
Obtaining $T_\mathrm{K}$ estimates in massive molecular datasets usually requires expensive observations of multiple transitions at different frequencies (e.g. CO ladders) and complex fitting procedures of thousands of spectra (e.g. NH$_3$). In most cases, this analysis is also complicated by the necessary combination with complex chemical and radiative transfer calculations in order to evaluate observational biases such as opacity, excitation, and line-of-sight effects. Moreover, the application of many of these methods requires the underlying assumption of local thermodynamic equilibrium (LTE; $T_\mathrm{K} = T_\mathrm{ex}\sim T_\mathrm{rot}$) and/or an effective dust-to-gas coupling ($T_\mathrm{K}=$~$T_\mathrm{dust}$), properties that are not necessary satisfied under most ISM conditions \citep[e.g.][]{GOL01}. These observational biases have usually limited the application of these techniques to some dedicated studies. In contrast, the detailed description of the temperature structure of the gas in molecular clouds remains largely unconstrained. 

Following one of the long-standing debates in astrochemistry \citep[e.g.][]{HER00}, this paper explores the use of the HCN-to-HNC line ratio as a new chemical thermometer of the molecular ISM. 
The favourable observational conditions and widespread detection of both HCN (cyanide) and HNC (isocyanide) isomers  make this observable a potential direct probe of the of the gas temperature within a wide range of scales and densities in the ISM. 
Using a new set of large-scale IRAM30m observations in the \object{Orion} A cloud (Sect.~\ref{sec:observations}),  we compare in this work the observed variations of the $I$(HCN)-to-$I$(HNC)  intensity ratio ($I$(HCN)/$I$(HNC)) of the J=1--0 lines
 with both independent temperature estimates 
and chemical models (Sect.~\ref{sec:results}). Our observational and model results demonstrate the strong sensitivity and systematic dependence of this $I$(HCN)/$I$(HNC)  with the gas kinetic temperature $T_\mathrm{K}$ between $\sim$~10 and 50~K. 
We calibrate this empirical correlation (Sect.~\ref{sec:Tk_calibration}) in order to obtain a large-scale temperature map of the \object{Orion A} cloud (Sect.~\ref{sec:application}). Applied to other star-forming regions and galactic surveys, our temperature estimates show an excellent correlation with independent dust temperature  measurements in a wide range of ISM environments within an optimal range between 15 and 40~K (Sect.~\ref{sec:universal}). Driven by the effective destruction of HNC, plus its isomerization into HCN, this temperature dependence could potentially explain the enhanced $I$(HCN)  intensities that are observed in star-forming galaxies (Sect.~\ref{sec:extragalactic}).


\section{Large-scale IRAM30m observations in Orion}\label{sec:observations}

\begin{figure*}
	\centering
	\includegraphics[width=\textwidth]{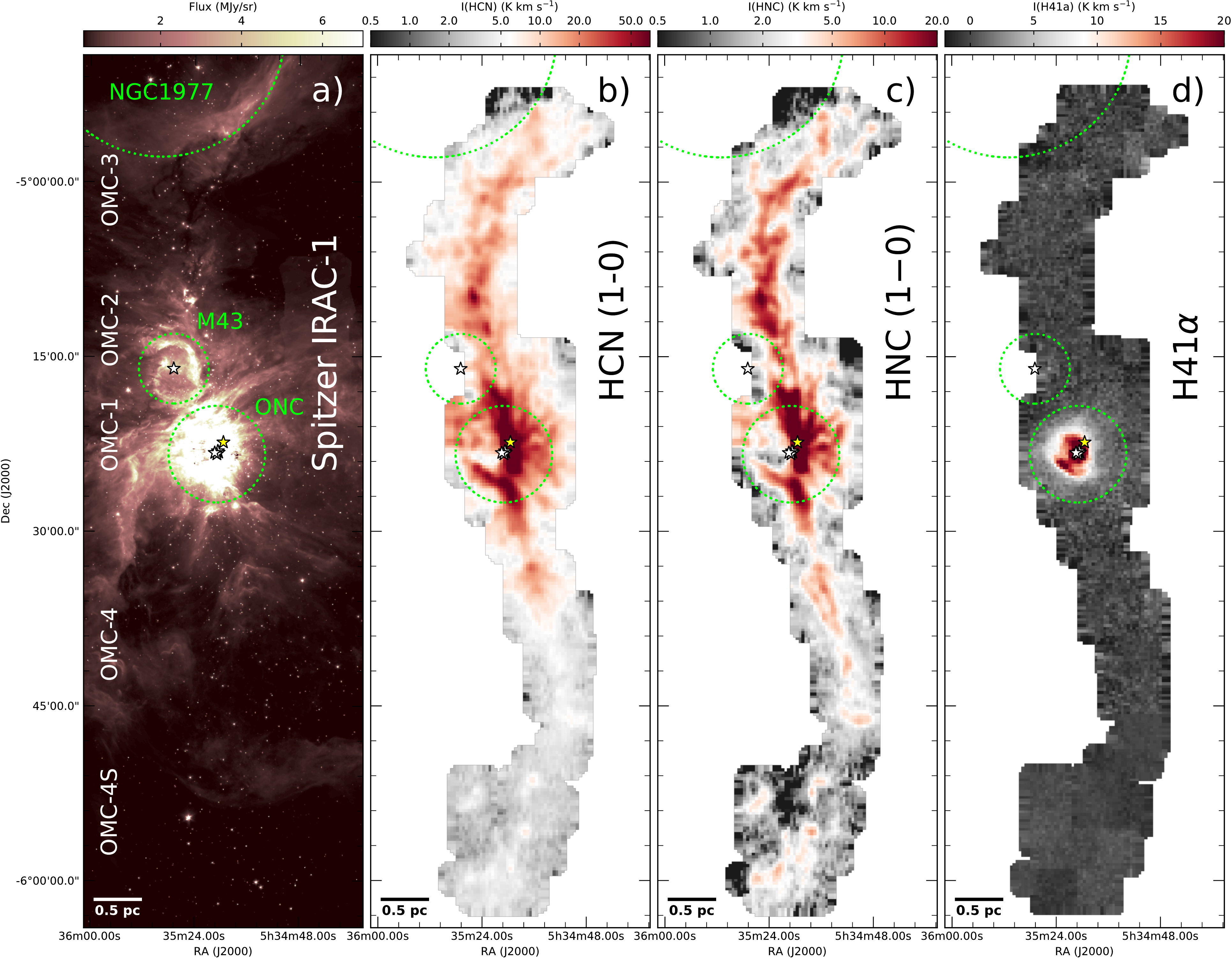}
	\caption{New  IRAM30m observations throughout the Orion ISF. From left to right: {\bf (a)} Spitzer IRAC-1 emission map \citep{MEG12}; {\bf (b)} HCN (J=1-0), {\bf (c)} HNC (J=1-0), and {\bf (d)} H41$\alpha$ line intensity (also known as total integrated intensity) maps (this work). We note that both HCN and HNC  intensity maps are presented in logarithmic scales because of the wide dynamic range in emission that these two species show. The different OMC 1-4 clouds, together with the different clusters and nebulosities, are indicated in the IRAC-1 image. For guidance, the extension of the ONC, M43, and NGC1977 regions (green dotted circles), as well the position of the Trapezium and NU Ori stars (white stars) plus the Orion BN source (yellow star), are also indicated in the different  IRAM30m maps. 
	}
	\label{fig:ISF_maps}
\end{figure*}

We systematically investigated the emission of the ground (J=1-0) transitions of the HCN \citep[$\nu=$~88.631 GHz, CDMS;][]{CDMS} and HNC \citep[$\nu=$~90.663 MHz, CDMS;][]{CDMS} isotopomers throughout the \object{Integral Shape Filament} (ISF) in Orion \citep{BAL87} using a new set of large-scale  IRAM30m molecular maps\footnote{
	This work is part of the ORION-4D project (PI: A. Hacar). See more information in \url{https://sites.google.com/site/orion4dproject} .}.
Our observations follow the main spine of this massive filament as it is delineated in the dust  continuum \citep{JOH99, LOM14} and dense tracers such as NH$_3$ \citep{FRI17} and N$_2$H$^+$ \citep{TAT08,HAC17a}. As illustrated in Fig.\ref{fig:ISF_maps}, our  observations cover a total area of $1.1 \times 0.14$~deg$^2$ in size, that is, $\sim8 \times 1 $~pc$^2$ at distance of 414~pc \citep{MEN07}, including the OMC 1-4S clouds, the entire Orion Nebula Cluster (\object{ONC}), the western half of the \object{M43} nebula, as well as the southernmost end of the \object{NGC 1977} cluster \citep[see][for a description of these regions]{PET08}.

The bulk of our  IRAM30m observations correspond to the central part of the ISF that extends throughout the OMC 1-4 clouds (proj.ID: 032-13), as has been presented by \citet{HAC17a}. Together with the N$_2$H$^+$ (1-0) line that was previously surveyed at high-velocity resolution, we simultaneously observed the HCN (1-0) and HNC (1-0) lines using the EMIR receiver connected to the FTS backend configured to a fine spectral resolution of 195~kHz, which corresponds to $\sim$~0.7~km~s$^{-1}$ at the frequency of these lines. In order to produce a large-scale mosaic, we combined different Nyquist-sampled on-the-fly (OTF) maps, typically of $200\times200$ arcsec$^2$ each, observed in position-switching (PSw) mode.  
Following standard procedures, we carried out atmospheric calibration every 15~min, and focus and pointing corrections every 1.5-2~hours. Three distinct positions throughout the cloud were systematically observed every 2 hours as reference spectra for cross-calibration between different runs \citep[see][for additional information]{HAC17a}.

We complemented our previous large-scale maps with additional observations of the OMC-4 South (OMC-4S) region (proj.ID: 034-16) carried out between August and November 2016 under average summer conditions with PWV=5mm. 
We used a spectroscopic EMIR + FTS setup and mapping strategy similar to our previous observing campaign, this time in frequency-switching mode (FSw). In addition to the corresponding standard pointing and focus measurements, the same reference spectra as were sampled in PSw mode throughout OMC-1 were re-observed in FSw mode. We found consistent intensities within the $\sim15\%$ calibration uncertainties of the telescope. 

An additional EMIR sideband was connected to an additional FTS unit in order to simultaneously survey the H41$\alpha$ recombination line ($\nu=$~92.034 GHz). The central part of this dataset was presented by \citet{GOI15} for the analysis of the [CII] and molecular emissions in the ONC. Similar to our previous molecular dataset, our last observing campaign extended these observations toward the OMC-4S region to create a large mosaic of the entire ISF. 

All of our  IRAM30m observations were reduced using the GILDAS/CLASS software
\footnote{
 \url{http://www.iram.fr/IRAMFR/GILDAS}
}.
Each individual dataset, with native resolution of $\sim$~27'', was independently combined and convolved into a common Nyquist-sampled grid with a final resolution of 30'' with a total $\sim$~8950 independent spectra for each of the lines in our survey. Because large velocity variations are observed within this cloud  \citep[e.g.][]{GOI15,HAC17a}, we applied local baseline corrections to each individual spectrum, for which we adapted each line window according to the specific target line and position in our maps.
We converted the observed line intensities, originally in antenna temperature units, into main beam temperature units using facility-provided efficiencies
\footnote{
	\url{http://www.iram.es/IRAMES/mainWiki/Iram30mEfficiencies}
}.
Finally, we obtained total integrated emission maps (i.e. $I(A)=\int T_{mb}(A)~dv$, in units of K~km~s$^{-1}$) between $[-20,20]$~km~s$^{-1}$ for the HCN (1-0) (i.e. including all hyperfine components), $[-10,15]$~km~s$^{-1}$ for HNC (1-0) transitions, and between $[-25,20]$~km~s$^{-1}$ for the H41$\alpha$.
These integration limits for HCN and HNC aim to capture the line intensities that originates at the cloud velocities. The choice of this velocity range deliberately excludes the emission of the high-velocity wings in the Orion BN/KL region. The line  intensities in this particular area should therefore be considered as lower limits\footnote{The integrated intensity maps for the HCN (1-0), HNC (1-0), and H41$\alpha$ lines presented in Figure~\ref{fig:ISF_maps} are available via CDS.}.

\section{Results}\label{sec:results}

\subsection{Observational correlation between HCN-to-HNC line ratio and the gas kinetic temperature}\label{sec:Tk_maps}

\begin{figure*}[ht!]
	\centering
	\includegraphics[width=\textwidth]{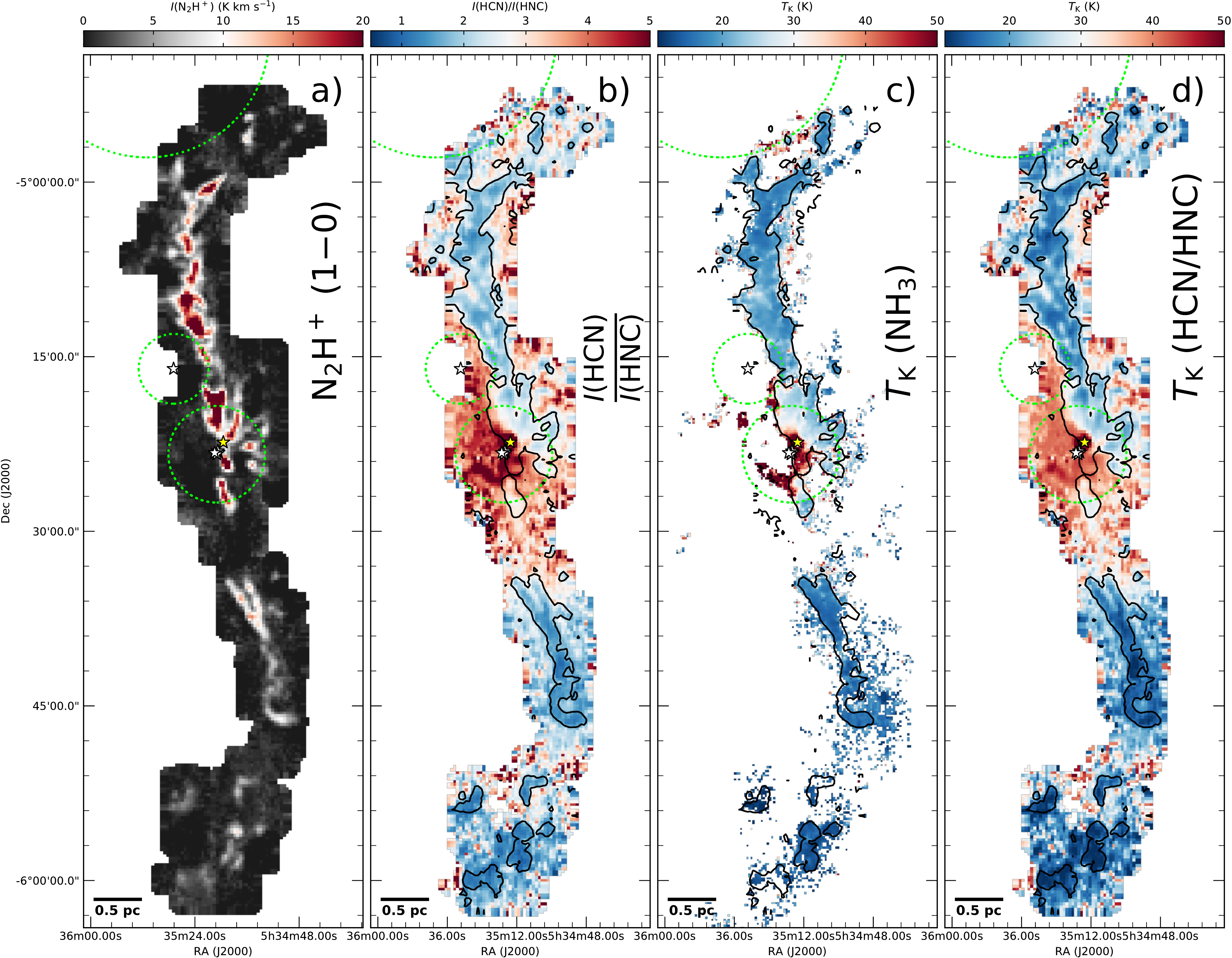}
	\caption{From left to right:
	{\bf (a)} N$_2$H$^+$ (1-0) integrated emission \citep[see also][]{HAC17a},
	{\bf (b)} $I\mathrm{(HCN)}$-to-$I\mathrm{(HNC)}$ line intensity ratio (see also Fig.~\ref{fig:ISF_maps}),
	{\bf (c)} gas kinetic temperature map derived using NH$_3$ measurements \citep{FRI17},
	and {\bf (d)} gas kinetic temperature map derived using the proposed $I\mathrm{(HCN)}/I\mathrm{(HNC)}$ as a temperature probe according to Eqs.~\ref{eq:Tk_HCN} and \ref{eq:Tk_HCN_high} (this work). For comparison, we indicate the intensity contour with $I$(N$_2$H$^+$)=1.5~K~km~s$^{-1}$ in panels (b-d). Circles and stars are similar to Fig.~\ref{fig:ISF_maps}.
	}
	\label{fig:ISF_Tkmaps}
\end{figure*}

We present the HCN and HNC (J=1--0) total integrated intensity maps (including all hyperfine components) throughout the ISF in Fig.\ref{fig:ISF_maps}.
As illustrated in panels (b-d), 
the emission maxima of these two species correspond to the central part of the ONC traced by H41$\alpha$ emission showing extended regions with integrated intensities above $I$(HCN)~$\ge$~50~K~km~s$^{-1}$ and $I$(HNC)~$\ge$~20~K~km~s$^{-1}$, respectively. Clearly recognizable at the centre of our images, the emission of both HCN and HNC isotopomers highlights most of the well-known OMC-1 molecular fingers \citep[e.g.][]{MAR90} and the Orion bar \citep[e.g.][]{TIE93}, which are seen in high contrast with respect to their local environment. Prominent emission in these two lines, with $I$(HCN)~,~$I$(HNC)~$>$~10~K~km~s$^{-1}$, also traces the northern part of the ISF towards the OMC-2 and OMC-3 clouds, showing multiple local peaks coincident with the position of several FIR sources within these regions (e.g. OMC-2 FIR-4). On the other hand, both HCN and HNC intensities progressively decrease below $I$(HCN)~,~$I$(HNC)~$<$~5~K~km~s$^{-1}$ towards the OMC-4 and OMC-4S clouds.  

The bright emission detected in the HCN and HNC (1-0) line maps reflects the large dynamic range of column densities that is traced by these two species.
While largely varying in intensity, both transitions are systematically detected in the vast majority of the positions surveyed by our IRAM30m observations. In particular, most of our HCN (99\%) and HNC (93\%) spectra show intensities above $I$(HCN)~,~$I$(HNC)~$\ge$~1~K~km~s$^{-1}$, that is, with a signal-to-noise ratio, S/N$>$~3, with respect to the typical $\sigma\sim$~0.3~K in our maps. Independent FIR {\it Herschel} measurements \citep{LOM14} indicate that these detections extend down to equivalent gas column densities of A$_V\sim$~3~mag within the limits of our molecular maps. Similar to previous observations of this region \citep{KAU17}, these detection thresholds indicate the presence and strong sensitivity of these HCN and HNC species not only to the high density gas but also to more extended and low column density material in this cloud \citep[see also][]{PET16,KAU17}.

While qualitatively similar in their overall distribution, we find large variations between the relative intensities of both HCN and HNC (1-0) transitions throughout the ISF. 
To illustrate this property, we present the total line intensity ratio $I\mathrm{(HCN)}/I\mathrm{(HNC)}$ of these two lines in Fig.~\ref{fig:ISF_Tkmaps}b.
An eye inspection of this figure indicates that  $I\mathrm{(HCN)}/I\mathrm{(HNC)}$ varies more than an order of magnitude throughout the areas that are surveyed in our maps. In agreement with previous results \citep{GOL86,SCH92,UNG97}, the largest differences in emission are found in the surroundings of the Orion BN/KL region showing $I\mathrm{(HCN)}/I\mathrm{(HNC)} >10$. This line ratio decreases towards values of $I\mathrm{(HCN)}/I\mathrm{(HNC)} \sim 2-3$ in regions such OMC-2, and more prominently, down to values of $I\mathrm{(HCN)}/I\mathrm{(HNC)} \sim 1$ in OMC-4. 

In addition to these regional differences, we note a systematic dependence of this $I\mathrm{(HCN)}/I\mathrm{(HNC)}$ as a function of column density. Particularly visible in the radial distribution of regions such as OMC-2, this reported line ratio varies from  values of $\sim2-4$ at the cloud edges towards values close to unity at the centre of the ISF. Although less prominent in dynamic range, a similar trend is also visible in regions like OMC-4, showing variations between $\sim3-1$. 

\begin{figure*}[ht!]
	\centering
	\includegraphics[width=0.7\linewidth]{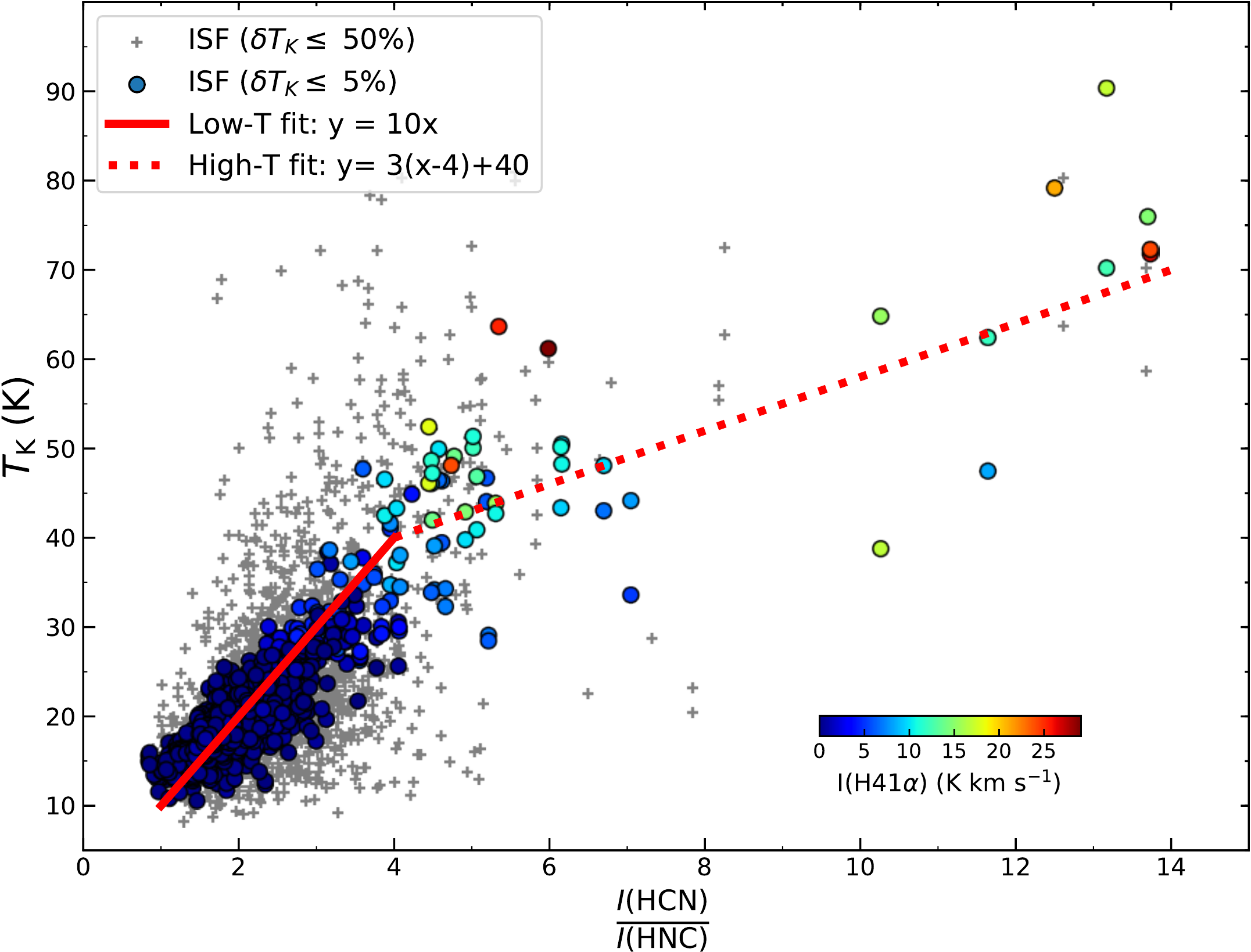}
	\caption{Correlation between the observed $I\mathrm{(HCN)}/I\mathrm{(HNC)}$ (this work) and measurements of the gas kinetic temperatures derived using NH$_3$ observations \citep{FRI17} with reliable temperature estimates (i.e. $\delta T_\mathrm{K}/T_\mathrm{K} \le 50$\%; grey crosses). These positions with good temperature estimates (i.e. $\delta T_\mathrm{K}/T_\mathrm{K} \le 5$\%) are colour-coded according to their total H41$\alpha$ intensity (see Fig.~\ref{fig:ISF_maps}). A systematic increase in the gas kinetic temperatures in Orion is clearly visible in positions with strong $I$(H41$\alpha$) emission, which denotes their proximity to the ONC. The empirical linear fit of each of the low- (solid red line) and high-(dotted red line) temperature regimes are indicated in the plot (see Sect.~\ref{sec:Tk_calibration}).
 	}
	\label{fig:ISF_TK}
\end{figure*}

Interestingly, we find an excellent correspondence between the variations of the $I\mathrm{(HCN)}/I\mathrm{(HNC)}$ ratio in our ISF data and the gas kinetic temperatures ($T_\mathrm{K}$) using NH$_3$ observations derived by the GBT-GAS survey \citep{FRI17}. As illustrated by the comparison of Fig.~\ref{fig:ISF_Tkmaps}~b and Fig.~\ref{fig:ISF_Tkmaps}~c, the lowest $I\mathrm{(HCN)}/I\mathrm{(HNC)}$ values are typically found at the coldest regions at the centre of clouds like OMC-2 or OMC-4. Conversely, higher $I\mathrm{(HCN)}/I\mathrm{(HNC)}$ values are shown at increasing gas temperatures  in regions such as OMC-1 and toward the cloud edges.

This dependence between $I\mathrm{(HCN)}/I\mathrm{(HNC)}$ and $T_\mathrm{K}$ becomes apparent in the point-to-point comparison displayed in Figure~\ref{fig:ISF_TK}. There, we include all the positions surveyed in our maps showing $I$(HCN)~,~$I$(HNC)~$\ge$~1~K~km~s$^{-1}$ and temperature estimates better than 50\%, that is, $(\delta T_K/T_K) \le 0.5$, according  to \citet{FRI17} (grey crosses). As a global trend, we identify a systematic increase of the $I\mathrm{(HCN)}/I\mathrm{(HNC)}$ ratio as function of temperature throughout the entire ISF, ranging from about $L\mathrm{(HCN)}/L\mathrm{(HNC)}  \sim 1$ at T$_K\le$~15~K to $L\mathrm{(HCN)}/L\mathrm{(HNC)}  \ge 5$ at T$_K\ge$~40~K. The distribution of these points indicates a typical dispersion of about $\pm$~5~K with respect to this average behaviour. 
The use of only those positions with high-quality temperature estimates, namely, better than 5\% (or $(\delta T_K/T_K) \le 0.05$) (blue solid points), significantly reduces this scatter, which suggests that a large fraction of this latter dispersion may be  produced by the uncertainties associated with these previous temperature measurements.

\subsection{Enhanced HCN-to-HNC abundance ratios at high gas temperatures}\label{sec:ratios}

The systematic variation in the HCN and HNC intensity ratio as a function of the gas kinetic temperature shown in Sect.~\ref{sec:Tk_maps} suggests a direct connection between the emission and thermal properties of the gas within the ISF. 
A priori, the variations in observed line ratios can potentially originate from the distinct excitation, opacity, and abundance of these two isomers.
In this section we examine which of these mechanisms are responsible of the observed $I\mathrm{(HCN)}/I\mathrm{(HNC)}$ variations.

Because the abundance of HCN is generally larger than that of HNC in regions like the ONC \citep[e.g.][]{GOL81}, the reported variations in the $I\mathrm{(HCN)}/I\mathrm{(HNC)}$ measurements can potentially be  explained by an increase in HCN (1-0) line opacity. 
Theoretical and observational results indicate an inverse correlation between the observed cloud temperatures as a function of the cloud depth.
If the HNC (1-0) line remained optically thin, the saturation of the HCN (1-0) line would then reduce the observed $I\mathrm{(HCN)}/I\mathrm{(HNC)}$ values in regions of increasing column densities, and therefore, decreasing temperatures.
We measured the HCN (1-0) opacities from the analysis of the hyperfine structure of all the spectra in our maps. Assuming a single-line component, we fitted each individual spectrum in our survey using the {\it hfs} method in CLASS assuming the hyperfine frequencies and relative intensities provided by the JPL database \citep{JPL} \footnote{
	For simplicity, we fitted all our spectra using a single-line component throughout the entire ISF. This assumption deliberately ignores the presence of more a complex kinematic substructure observed at interferometric resolutions in regions like OMC-1 and OMC-2 \citep{HAC18}. Previous single-dish studies indicate that most of this complexity is smoothed out at the resolution of our  IRAM30m observations \citep{HAC17a} potentially affecting our opacity estimates. These resolution effects are assumed to have a minor statistical impact in our large-scale analysis. Nevertheless, these caveats should be considered on the interpretation of individual spectra.
}.
A total of 7810 HCN (J~=~1-0) spectra were fitted with S/N$\ge$~3. Of these, 83\% are found to be optically thin, showing opacities of their central hyperfine component with $\tau$(F~=~2--1)~$\le$~1. These optically thick spectra, 98\% of which show $\tau$(F~=~2--1)~=1-3, are primarily concentrated in high column density areas within the OMC-4 and OMC-4S clouds. In contrast, most of the OMC-1, OMC-2, and OMC-3 spectra are found to be optically thin with opacities as low as $\tau$(F~=~2--1)~$\lesssim$~0.1.
Nonetheless, we find hyperfine anomalies \citep{WALM82} in many of our HCN spectra around the ONC making their opacity estimates uncertain.
In the absence of additional measurements (e.g. H$^{13}$CN), radiative transfer calculations using RADEX \citep{RADEX} demonstrate that the expected opacity variations can be responsible for changes of the $I$(HCN)/$I$(HNC) ratio up to a factor of $\sim$~3 (see also Appendix~\ref{sec:appendix2}).
While likely affecting some spectra in our maps,
these estimates allow us to rule out opacity as the main driver of the global variations in the $I\mathrm{(HCN)}/I\mathrm{(HNC)}$ values observed throughout the ISF.

With minor opacity effects, the similar frequencies, dipole moments, and excitation conditions of the HCN and HNC (1-0) transitions make their intensity ratio a good proxy of the relative column densities of these two species \citep{GOL86}. 
In this sense, the results obtained in Sect.~\ref{sec:observations} would suggest a systematic variation of the relative abundances of the HCN and HNC molecules with respect to the gas kinetic temperature.
We have quantified these abundance variations using RADEX. 
A series of simple tests with constant X(HCN) and X(HNC) abundances confirm the weak dependence of the observed $I\mathrm{(HCN)}/I\mathrm{(HNC)}$ values as function of temperature (with changes of a factor $\lesssim$~2) for a given isomer ratio.
The observed variation ratios also exceed the uncertainties produced by the different collisional rate coefficients between HCN and HNC \citep[$\sim$5 at low temperatures,][]{HER17}.
Instead, only large isomeric abundance differences, of more than and order of magnitude in relative abundance (i.e. $X\mathrm{(HCN)}/X\mathrm{(HNC)} =1-20$), are able to reproduce the observed variations of this line intensity ratio as function of $T_\mathrm{K}$ shown in Fig.~\ref{fig:ISF_TK}.

\subsection{HNC destruction mechanisms: low-energy barriers}\label{sec:Ebarrier}

Understanding the origin of the observed temperature dependence of the $X\mathrm{(HCN)}/X\mathrm{(HNC)}$ abundance ratio in molecular clouds is an old question in astrochemistry studies \citep[e.g.][]{WOO78,HER00}.
Large $X\mathrm{(HCN)}/X\mathrm{(HNC)}$ abundance variations, by more than an order of magnitude, are regularly reported both within and between clouds as a function of temperature \citep{GOL81,IRV84}.
Extreme ratios of $X\mathrm{(HCN)}/X\mathrm{(HNC)} >30$ have traditionally been found at high temperatures in regions such as the ONC \citep{SCH92}. 
In contrast, much lower abundance ratios of $X\mathrm{(HCN)}/X\mathrm{(HNC)} \sim 0.7$ are typically observed in dense cores at lower temperatures \citep{HAR89,HIR98}. 
Compared to these previous results, the large dynamic range shown in Fig.~\ref{fig:ISF_TK} indicates a smooth connection between these warm and cold environments.

\begin{figure*}[ht!]
	\centering
	\includegraphics[width=0.60\textwidth]{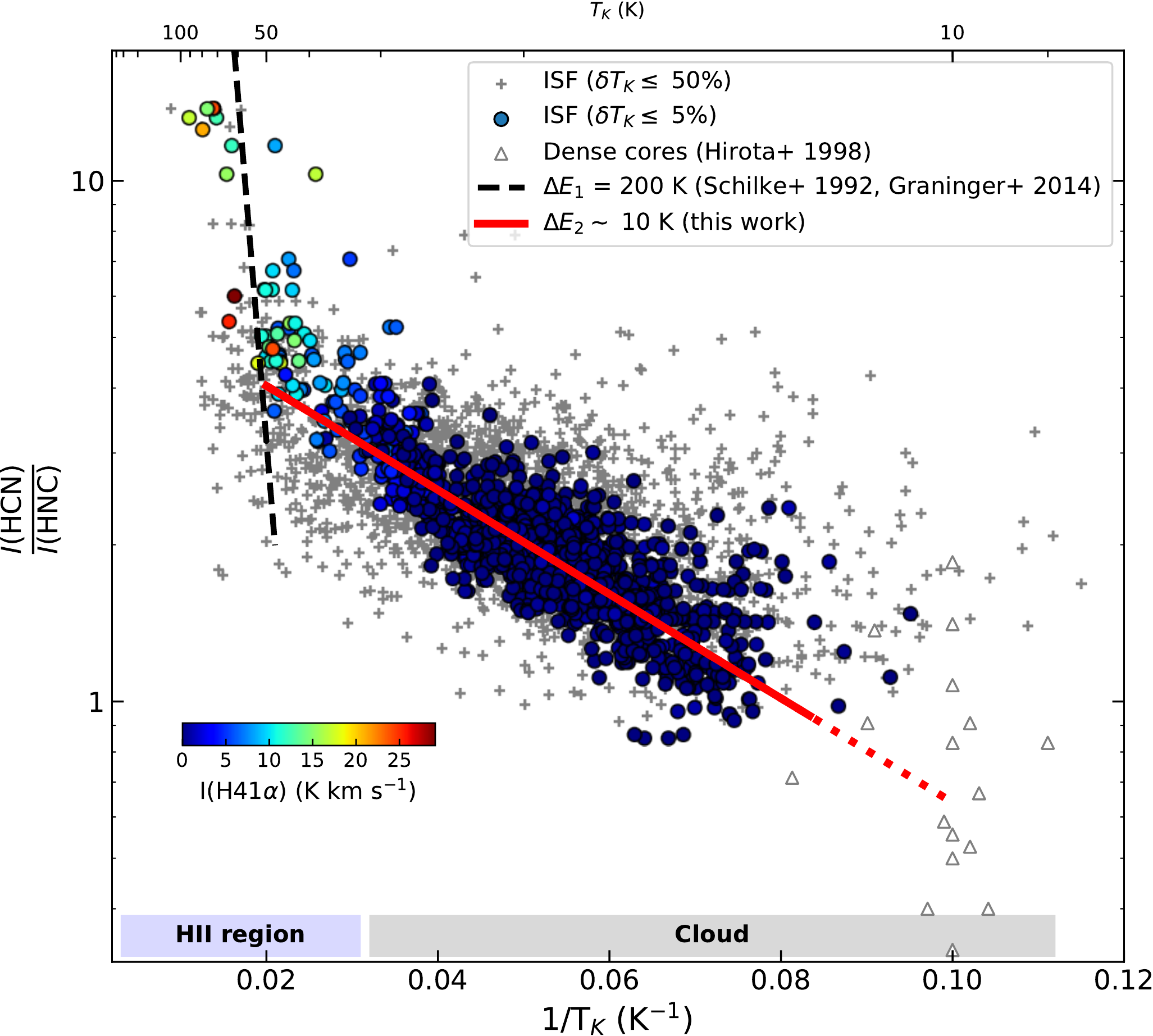}\vskip0.4cm
	\includegraphics[width=0.60\textwidth]{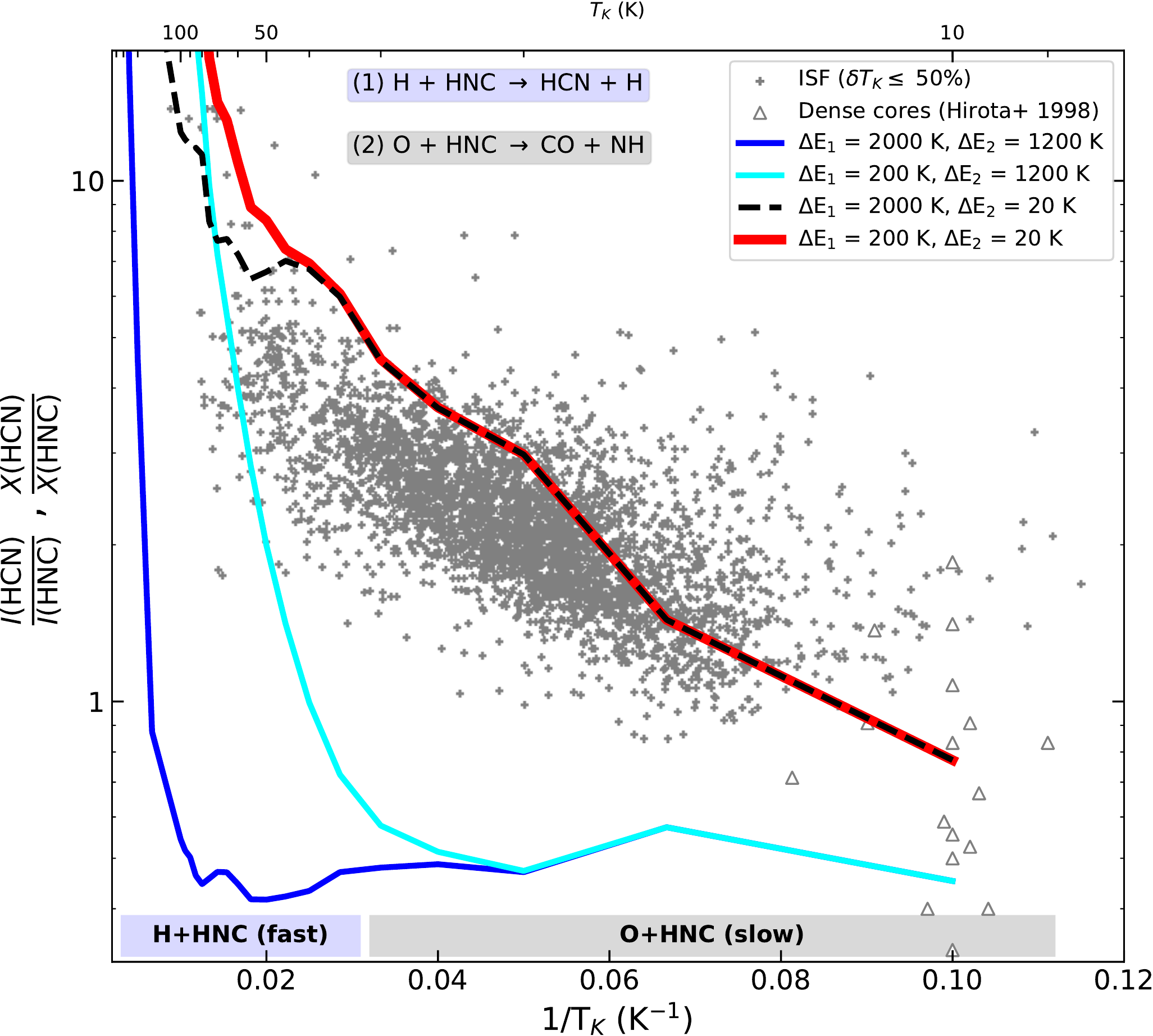}
	\caption{
	{\bf (Upper panel)}  Observed dependence of the $I\mathrm{(HCN)}/I\mathrm{(HNC)}$  intensity ratio as function of 1/$T_\mathrm{K}$ in the ISF (grey crosses; this work) in comparison with classical estimates from dense core surveys \citep[grey triangles;][]{HIR98}. The positions with good temperature estimates (i.e., $\delta T_\mathrm{K}/T_\mathrm{K} \le 5$\%) are colour-coded according to their total H41$\alpha$ intensity (see Fig.~\ref{fig:ISF_maps}). The temperature dependence of the $I\mathrm{(HCN)}/I\mathrm{(HNC)}$ expected for an energy barrier of $\Delta E_1=$~200~K for reaction \ref{R1} is indicated by the black dashed line \citep{SCH92,GRA14}. Similarly, the newly proposed low energy barrier for $\Delta E_2=$~10~K for reaction \ref{R2} is shown in red. 
	{\bf (Lower panel)} $X\mathrm{(HCN)}/X\mathrm{(HNC)}$ abundance variations predicted by our models (1-4) with different energy barriers (see also text) compared to the observed intensity variations in cores (grey triangles) and throughout the ISF (grey crosses). Our model results illustrate the different dominant roles of reactions (\ref{R1}) and (\ref{R2}) determining the temperature structure of the ONC (blue area) and the rest of the cloud (grey area).
	The detailed equivalence between the observed intensity ratios and the corresponding abundance ratios in our models are discussed in Appendix~\ref{sec:appendix1}.
}
	\label{fig:EBarrier}
\end{figure*}

Different theoretical studies have suggested that the observed $X\mathrm{(HCN)}/X\mathrm{(HNC)}$ variations could be chemically controlled.
Laboratory experiments and chemical models agree upon the dissociative recombination of HCNH$^+$ as the main the formation pathway of HCN and HNC molecules, forming both isomers with an approximately branching ratio of 1:1 at T$_K<$~200~K \citep[e.g.][]{HER00}.
With the same initial abundances, the observed isomer variations are therefore regulated by their destruction mechanisms.
In particular, two neutral-neutral reactions have been proposed to be responsible for the observed temperature variations of the $X\mathrm{(HCN)}$-to-$X\mathrm{(HNC)}$ abundance ratios ($X\mathrm{(HCN)}/X\mathrm{(HNC)}$) in the ISM: 
\begin{equation}\label{R1}
\mathrm{HNC} + \mathrm{H} \rightarrow \mathrm{HCN} + \mathrm{H}
\end{equation}
and
\begin{equation}\label{R2}
\mathrm{HNC} + \mathrm{O} \rightarrow \mathrm{NH} + \mathrm{CO}
\end{equation}
Classical ab initio calculations estimate energy barriers of $\Delta E_1=$~1200~K and $\Delta E_2=$~2000~K for reactions (\ref{R1}) and (\ref{R2}), respectively \citep[see][for a full discussion]{GRA14}.
Recent chemical models demonstrate, however, that the observed $X\mathrm{(HCN)}/X\mathrm{(HNC)}$ variations in the vicinity of the ONC can be explained if the HNC+H reaction (\ref{R1}) possesses an energy barrier of $\Delta E_1=$~200~K, that is, approximately an order of magnitude lower than previously estimated  \citep{GRA14}. 
A similarly low-energy barrier has been proposed for the HNC+O reaction (\ref{R2}) \citep{JIN15}. Nevertheless, no observational nor theoretical work has quantified this latter energy barrier in detail to date. 

In Figure~\ref{fig:EBarrier} (upper panel) we display the observed variations of the $I\mathrm{(HCN)}$-to-$I\mathrm{(HNC)}$  intensity ratio in Orion, this time as a function of $1/T_{\mathrm{K}}$ (grey crosses). We colour-code those positions with accurate temperature estimates (with $\delta T_{\mathrm{K}}\le 5$\%) according to their H41$\alpha$ intensity in Fig.~\ref{fig:ISF_maps} as an indication of their proximity to the ONC.
Two separate regimes can clearly be distinguished in this plot. At $T_{\mathrm{K}}\gtrsim 50$~K, those positions directly exposed to the HII nebula (i.e. showing high values of H41$\alpha$ emission) show a rapid variation in $I\mathrm{(HCN)}/I\mathrm{(HNC)}$ as a function of $T_{\mathrm{K}}$. On the other hand, the cloud positions without significant H41$\alpha$ emission present a much shallower dependence for temperatures $T_{\mathrm{K}}\sim[10,50]$~K. Interestingly, this trend seems to continue down to the typical initial branching ratios ($\sim [0.5,1.0]$) found in dense cores \citep[grey triangles;][]{HIR98}. These distinct slopes suggest that two independent mechanisms control the observed $I\mathrm{(HCN)}/I\mathrm{(HNC)}$ variations with a surprisingly low dispersion (see the narrow spread of our measurement for a given $1/T_{\mathrm{K}}$ value in our plot).

Compared to Fig.~\ref{fig:ISF_TK}, the use of $1/T_{\mathrm{K}}$ units in Figure~\ref{fig:EBarrier} allows us to directly estimate the potential impact of different energy barriers for reactions (\ref{R1}) and (\ref{R2}).
Assuming that $\left(\frac{I\mathrm{(HCN)}}{I\mathrm{(HNC)}}\right) = \left(\frac{\mathrm{X(HCN)}}{\mathrm{X(HNC)}}\right)$  (see Appendix~\ref{sec:appendix2} for a full discussion), the value of $\Delta E_i$ can be directly obtained from the linear fit of $\left(\frac{\mathrm{X(HCN)}}{\mathrm{X(HNC)}}\right)=A\times \mathrm{exp}\left(\frac{-\Delta E_i}{T_{\mathrm{K}}}\right)$, that is, $\mathrm{log}_{10}\left(\frac{I\mathrm{(HCN)}}{I\mathrm{(HNC)}}\right)\propto \frac{-\Delta E_i}{T_{\mathrm{K}}}$ under this representation. As denoted in this figure (black dashed line), our observations reproduce the previously proposed steep dependence of the $I\mathrm{(HCN)}/I\mathrm{(HNC)}$ ratio at the high temperatures found within the ONC region using $\Delta E_1=$~200~K \citep[black dashed line][]{SCH92,GRA14}. These results are expected because the HNC~+~H reaction (\ref{R1}) is favoured by the large amount of free H atoms that are generated within the HII region. 

On the other hand, the shallower dependence shown in Fig.~\ref{fig:EBarrier} (upper panel) at low temperatures suggests a much lower energy barrier for the corresponding HNC~+~ O reaction (\ref{R2}). A series of manual fits to our data indicates an energy barrier for this latter reaction of about $\Delta E_2\sim$~10~K (red line). If confirmed, the chemical destruction of HNC via HNC+O reaction could potentially dominate the observed $I\mathrm{(HCN)}/I\mathrm{(HNC)}$ intensity variations at $T_{\mathrm{K}}< 50$~K within clouds.

We quantified the energy barriers in reactions (\ref{R1}) and (\ref{R2}) using a grid of standard chemical models.
Our chemical models use the code and network from \citet{BOS18}. The chemical network is based on gas-phase reactions  from the \textsc{Rate12} network from the UMIST Database for Astrochemistry
\footnote{Grain surface reactions from the Ohio State University (OSU) network:  \url{http://faculty.virginia.edu/ericherb/research.html} }
 \footnote{\url{http://www.udfa.net}} \citep{GAR08, MCE13, WAL15}. Binding energies where taken from \citet{PEN17}. The reaction rate coefficient for reaction (1) was replaced, and reaction (2) was added to the network, both with the values from \citet{GRA14}.

For our study, we created a set of four model grids (models 1-4) using different combinations of $(\Delta E_1,\Delta E_2)$ values (see below) and tracked their evolution with time.
Using the same chemical network, we aim to isolate the effects of distinct energy barriers on the observed isomeric fractionation as function of temperature.
We evaluated the corresponding $X\mathrm{(HCN)}$-to-$X\mathrm{(HNC)}$ abundance ratio in these models at temperatures between $T_{\mathrm{K}}=[10,1000]$~K sampled with 20 points per dex in log-scale.
In all cases, we assumed a typical evolutionary timescale of $\tau=$~0.5~Myr, a density of $n(\mathrm{H}_2)=10^4$~cm$^{-3}$, and a standard cosmic ionization rate of $\zeta=10^{-17}$~s$^{-1}$. These fiducial parameters are meant to capture the overall dependence of the HCN/HNC abundance variations under standard ISM conditions (see below). For completion, we describe the variations with respect to these fiducial models as a function of time, density, and cosmic-ray ionization in Appendix~\ref{sec:appendix1}.

We compare the $X\mathrm{(HCN)}/X\mathrm{(HNC)}$ abundances predicted by our chemical models 1-4 with the observed $I\mathrm{(HCN)}/I\mathrm{(HNC)}$  intensity variations in Orion (grey crosses) in Figure~\ref{fig:EBarrier} (lower panel; colour lines). The properties of these models can be summarized as follows:
\begin{itemize}
	\item  Model 1 uses the classical energy barriers estimated by ab initio calculations: $\Delta E_1=2000$~K and $\Delta E_2=1200$~K (dark blue line). With reactions (\ref{R1}) and (\ref{R2}) both inhibited by their high-energy barriers, model 1 shows no abundance variation respect to the original HCN/HNC branching value at the temperatures surveyed in Orion (i.e., $T_{\mathrm{K}}\ll \Delta E_i$), on contrast to our observations. 
	
	\item Model 2 assumes a $\Delta E_1=200$~K but keeps $\Delta E_2=1200$~K (cyan line). In close agreement to the results proposed by \citet{GRA14}, the effects of the low-energy barrier in the H~+~HNC reaction explain the rapid increase of $X\mathrm{(HCN)}/X\mathrm{(HNC)}$ that is observed at high temperatures. In this model 2, however, this reaction (\ref{R1}) alone fails to reproduce the observed temperature dependence at cloud temperatures of $T_{\mathrm{K}}\lesssim 50$~K. 

	\item Model 3 quantifies the impact of a low-energy barrier for the O~+~HNC reaction (\ref{R2}) assuming a $\Delta E_2=20$~K in the absence of reaction (\ref{R1}) suppressed by a $\Delta E_1=2000$~K (dashed black line). These calculations demonstrate the dominant role of reaction (\ref{R2}) at low temperatures. As predicted by our linear fits (see above), model 3 reproduces the smooth increase in $X\mathrm{(HCN)}/X\mathrm{(HNC)}$ with temperatures within 10~K~$\lesssim T_{\mathrm{K}}< 50$~K. In contrast, this model underestimates the much stronger temperature dependence that is observed at $T_{\mathrm{K}}> 50$~K in HII regions like the ONC.
	
	\item Finally, model 4 explores the combined effects of low-energy barriers in reaction (\ref{R1}; Model 2) and reaction (\ref{R2}; Model 3) adopting barriers with $\Delta E_1=200$~K and $\Delta E_2=20$~K, respectively (solid red line). In remarkably close agreement with our observations, model 4 simultaneously reproduces both low- and high-temperature variations in the $I\mathrm{(HCN)}/I\mathrm{(HNC)}$ in Orion. Moreover, it connects these results with the reported abundance ratios in dense cores. 

\end{itemize}

Our chemical models demonstrate the relative importance of the different HNC destruction mechanisms with low-energy barriers determining the observed $I\mathrm{(HCN)}/I\mathrm{(HNC)}$ variations in clouds like Orion. In agreement with previous results \citep{SCH92,GRA14}, warm and irradiated (HII) regions such as the ONC are largely dominated by the isomerization of HNC into HCN via H~+~HNC$\rightarrow$~HCN~+~H. Outside these active regions, our models suggest that the additional destruction of HNC via O~+~HNC~$\rightarrow$~NH~+~CO reaction is the most likely driver of the reported $I\mathrm{(HCN)}/I\mathrm{(HNC)}$ variations at lukewarm temperatures. 

The observational and model results both consistently demonstrate that reactions (\ref{R1}) and (\ref{R2}) have much lower activation energies  than previously estimated. Our Orion data confirm the previously proposed $\Delta E_1=200$~K barrier for the H~+~HNC reaction \citep{GRA14}. In addition, the large dynamic range of our data at intermediate temperatures indicates an almost barrier-less activation energy for the O~+~HNC reaction. Determining its precise value, however, is made difficult by the apparent scatter ($\sim$0.2~dex) in our data and the potential observational biases on the comparison between intensity and abundance ratios (see an extensive discussion of these comparisons in Appendix~\ref{sec:appendix2}).
Nonetheless, the unprecedented dynamic range of our data suggests an observational constraint on the energy barrier for reaction (\ref{R2}) of approximately $\Delta E_2\sim20$~K. Additional characterization of multiple HCN and HNC isotopologues, together with dedicated chemical models, are therefore needed to accurately quantify this barrier \citep[e.g. see][]{SCH92}. 

\subsection{HCN/HNC abundance variations: evolution and steady state}

\begin{figure}
	\centering
	\includegraphics[width=\linewidth]{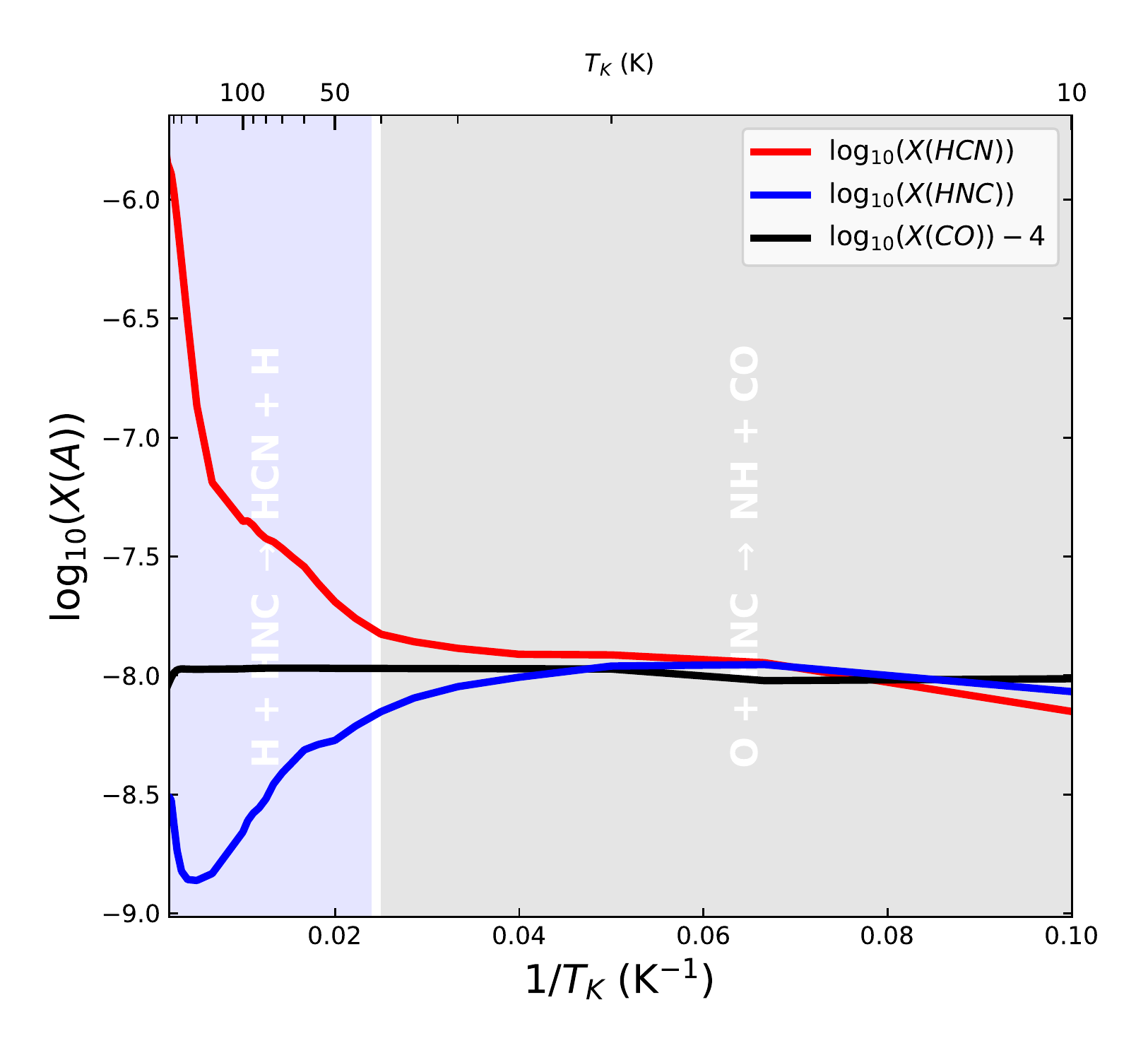}
	\caption{Temperature variations of the absolute abundance for HCN (red), HNC (blue), and $^{12}$CO (black) predicted by our model 4. The range of influence of the HNC~+~H (reaction \ref{R1})  and HNC~+~O (reaction~\ref{R2}) destruction mechanisms is indicated in blue and grey coloured areas, respectively.
	}
	\label{fig:AbsAbun}
\end{figure}

Our chemical models allow us to explore in high detail the origin of the observed temperature dependence of the $X\mathrm{(HCN)}/X\mathrm{(HNC)}$ found in Sect.~\ref{sec:Ebarrier}. 
Figure~\ref{fig:AbsAbun} shows the individual gas-phase abundances of HCN (red) and HNC (blue) with respect to H$_2$ (i.e. X(A)=X(A)/X(H$_2$)) as a function of temperature obtained in our previous model 4 (see Fig.~\ref{fig:EBarrier}, lower panel). At $T_{\mathrm{K}}=$~10~K (or $1/T_{\mathrm{K}}=0.1$~K$^{-1}$), both isomers show similar absolute abundances of X(HCN)$\sim$X(HNC)$= 10^{-8}$, as expected from the initial branching. 
The abundance variations of these two species are compared to the relative amount of CO (black), a reference molecule with almost constant abundance of X(CO)$\sim 10^{-4}$ at all temperatures in our models. The changes in the HCN and HNC abundances clearly illustrate the relative influence of reactions (\ref{R1}) and (\ref{R2}) in explaining the two separated temperature regimes observed in Fig.~\ref{fig:EBarrier}. 
At low $T_{\mathrm{K}}$ (or high $1/T_{\mathrm{K}}$), the smooth increase in the observed $X\mathrm{(HCN)}/X\mathrm{(HNC)}$ values is dominated by the reduction of the HNC abundance, up to 0.25 dex, after the recombination of this molecule with O in reaction (\ref{R2}). This selective destruction of HNC is exacerbated after the activation of reaction (\ref{R1}) and the subsequent production of HCN at higher temperatures. Together, the resulting $X\mathrm{(HCN)}/X\mathrm{(HNC)}$ abundance ratio rapidly increases by more than one order of magnitude at $T_{\mathrm{K}}>40$~K (or $1/T_{\mathrm{K}}<0.025$~K$^{-1}$).
 
During the early evolution of our models the destruction of HNC also proceeds at different speeds within each of the two temperature regimes identified before.
Because the large reservoir of free H atoms, the isomerization of HNC via reaction (\ref{R1})  occurs on short timescales of $\tau\sim 0.1$~Myr and rapidly dominates the destruction of this molecule at high temperatures. On the other hand, the recombination of O~+~HNC depends largely on the more limited atomic O reservoir that is regulated by the combined effect of gas density and cosmic-ray ionization (see Appendix~\ref{sec:appendix1}). Under the normal ISM conditions described in our model 4, reaction (\ref{R2}) proceeds at slower rates and becomes relevant only at $\tau > 0.3$~Myr. 

Despite their different initial speeds, the two HNC destruction reactions reach steady state at later times in our models.
For a given set of energy barriers, and independently of their density and cosmic-ray ionization, the observed $X\mathrm{(HCN)}/X\mathrm{(HNC)}$ ratio in our simulations follows an almost identical temperature dependence at timescales $\tau \ge 0.5$~Myr. This functional dependence remains unaltered in our models until the end of our simulations at $\tau = 3$~Myr (see Appendix~\ref{sec:appendix1} for a discussion). These results suggest a (pseudo-)equilibrium between the formation and destruction mechanisms of the HCN and HNC isomers. This stable configuration could explain the low dispersion observed in the $\gtrsim$~1~Myr old gas in Orion.


\section{A new chemical thermometer for the ISM}\label{sec:Tk_calibration}

In Sect.~\ref{sec:ratios} and \ref{sec:Ebarrier} we investigated the origin of the large variations of the observed $I\mathrm{(HCN)}$-to-$I\mathrm{(HNC)}$ intensity ratios for their J=1--0 lines in Orion. As illustrated by our models, these intensity differences correspond to actual changes on the relative abundances of both HCN  and HNC isomers due to the selective destruction of HNC.
These abundance variations are independent of the gas density, and after short timescales ($\sim 0.3$~Myr), they become invariant with time. The systematic dependence shown in our  data indicates that the absolute value of the reported $I\mathrm{(HCN)}/I\mathrm{(HNC)}$ ratios is primarily determined by the temperature dependence of reactions~\ref{R1} and \ref{R2}. 
In this section, we discuss how these observational variations can be employed as a direct measurement of the gas kinetic temperatures of the molecular gas in the ISM.

\subsection{Calibration: empirical correlation between HCN/HNC and $T_\mathrm{K}$}

As shown by the first-order linear fits in Fig.~\ref{fig:ISF_TK} (solid and dashed red lines), the correlation between the total (i.e. including all hyperfine components) observed $I\mathrm{(HCN)}/I\mathrm{(HNC)}$ ratios and gas kinetic temperatures $T_\mathrm{K}$ can be described by a two-part linear function following:
\begin{equation}\label{eq:Tk_HCN}
T_\mathrm{K}[\mathrm{K}]=10\times \left[\frac{I\mathrm{(HCN)}}{I\mathrm{(HNC)}}\right] \quad \mathrm{if}  \quad \left(\frac{I\mathrm{(HCN)}}{I\mathrm{(HNC)}}\right) \le 4
\end{equation} and
\begin{equation}\label{eq:Tk_HCN_high}
 T_\mathrm{K}[\mathrm{K}] =3\times \left[\frac{I\mathrm{(HCN)}}{I\mathrm{(HNC)}}-4\right]+40 \quad \mathrm{if} \quad \left(\frac{I\mathrm{(HCN)}}{I\mathrm{(HNC)}}\right) > 4
\end{equation}
At low-intensity ratios $1< I\mathrm{(HCN)}/L\mathrm{(HNC)} \le 4$, the goodness of our fit demonstrates that Eq.~\ref{eq:Tk_HCN} accurately predicts the thermal gas conditions of the gas at lukewarm temperatures with errors of $\Delta T_\mathrm{K}\lesssim$~5~K.
These uncertainties seem to grow at $I\mathrm{(HCN)}/I\mathrm{(HNC)} \lesssim 1$, describing low-temperature regimes of $T_\mathrm{K}\lesssim $~10~K, which are likely due to the combination of excitation and opacity effects.
On the other hand, Eq.~\ref{eq:Tk_HCN_high} is suggested for $I\mathrm{(HCN)}/I\mathrm{(HNC)} > 4$ in order to obtain rough estimates of higher gas temperatures, despite its relatively larger uncertainties with $\Delta T_\mathrm{K}>$~10~K.  Part of these latter discrepancies could be generated by the limited sensitivity of the low rotational transitions of NH$_3$ when gas temperatures above T$_K>50$~K are traced \citep{TAN17}.

The empirical calibration of Equations \ref{eq:Tk_HCN} and \ref{eq:Tk_HCN_high} introduces a simple method for estimating the gas kinetic temperatures. 
Their large abundances and favourable excitation conditions make HCN and HNC two of the most frequently observed molecular tracers in current line surveys that are easily detectable in local and extragalactic studies alike \citep[e.g.][]{PET16}. Because of their proximity in frequency, transitions such as HCN (1-0) and HNC (1-0) can also be simultaneously observed using standard broad-band receivers (e.g. Fig.~\ref{fig:ISF_maps}).  
Our results demonstrate the direct use of the $I\mathrm{(HCN)}/I\mathrm{(HNC)}$ as a proxy of the gas kinetic temperature within clouds in a wide range of physical conditions (see also Sects.~\ref{sec:application} and \ref{sec:universal}). 
Readily obtained at large-scales, the proven sensitivity of this straightforward observable opens a novel window on the study of the thermal gas properties of the ISM without the need of any chemical or radiative transfer model. Moreover, and unlike the dust emission, this method offers the possibility of studying the temperature structure of the different gas components that are superposed along the line-of-sight when these are resolved in velocity.
In summary, we propose the use of the empirical correlations described by Eqs.~\ref{eq:Tk_HCN} and \ref{eq:Tk_HCN_high} as a new chemical thermometer of the molecular gas in the ISM with an optimal working range between 15~K~$\lesssim  T_\mathrm{K} \le $~40~K.

\subsection{Application: gas temperature map of the Orion ISF}\label{sec:application}

We have tested the use of the proposed $I\mathrm{(HCN)}/I\mathrm{(HNC)}$ as a temperature probe from the analysis of our own data in Orion. 
For each position in our maps with detected emission in each of the HCN and HNC isomers $I$(HCN)~,~$I$(HNC)$\ge$~1.0~K~km~s$^{-1}$, we calculated the gas kinetic temperature $T_\mathrm{K}$(HCN/HNC) according to Equations \ref{eq:Tk_HCN} and \ref{eq:Tk_HCN_high}.
We present the resulting temperature maps in Figure~\ref{fig:ISF_Tkmaps}d. An inspection of Fig.~\ref{fig:ISF_Tkmaps} reveals the potential of this new technique for the study of the thermal structure of clouds such as Orion. As illustrated by this figure, the observed  $I\mathrm{(HCN)}/I\mathrm{(HNC)}$ variations capture the well-known large temperature differences between the warm ONC region and the much colder surrounding cloud material. In particular, our maps show the systematic temperature differences throughout the ISF when comparing the warm OMC-1/2/3 regions with the colder OMC-4/4S clouds.
Although calculated on a point-to-point basis, the observed HCN/HNC derived temperatures present a smooth spatial variation that denotes the good behaviour of these measurements across our maps. 

We have quantified the performance of our HCN/HNC temperature estimates (Fig.~\ref{fig:ISF_Tkmaps}d) in comparison to the previously derived temperatures maps obtained from ammonia observations \citep{FRI17} (Fig.~\ref{fig:ISF_Tkmaps}c).
Overall, our HCN/HNC estimates partially resolve the rapid increase of the gas kinetic temperature in the OMC-1 region towards the Trapezium stars and the ONC, showing consistent gas temperatures $T_\mathrm{K} > $~30~K. Nonetheless, our measurements underestimate the extreme temperatures found around the Orion BN/KL region as well as in the Orion Bar detected in warm NH$_3$. This is not surprising given the larger uncertainties of our method at the high temperature regimes described by Eq.\ref{eq:Tk_HCN_high} above $T_\mathrm{K} > $~40~K (see Sect.\ref{sec:Tk_calibration}). Outside these problematic areas, our HCN/HNC temperature measurements show a remarkably close agreement with previous NH$_3$ estimates. In particular, we find an excellent correlation between the absolute and relative variations of these two temperature  estimates in dense and cold regions such as cores and filaments with significant N$_2$H$^+$ emission \citep[enclosed by a black contour in our maps in Fig.~\ref{fig:ISF_Tkmaps}; see][]{HAC17a,HAC18}. 
 
In addition, the detection of the HCN and HNC isomers at large scales allows us to investigate the temperature structure of the gas at low column densities. As shown in Fig.~\ref{fig:ISF_Tkmaps}d, our $T_\mathrm{K}$(HCN/HNC) measurements extend far beyond regions with N$_2$H$^+$ or NH$_3$ detections. In particular, the significant increase in temperature measurements towards low-density regions such as OMC-4/4S is noteworthy.
Interestingly, our estimates show a rapid increase in the gas temperature to values above $T_\mathrm{K} \gtrsim $~30~K in regions with no detected dense gas. 

\subsection{Validation: comparison with other temperature estimates in Orion A}\label{ap:comparison}

\begin{figure*}
	\centering
	\includegraphics[width=1.0\linewidth]{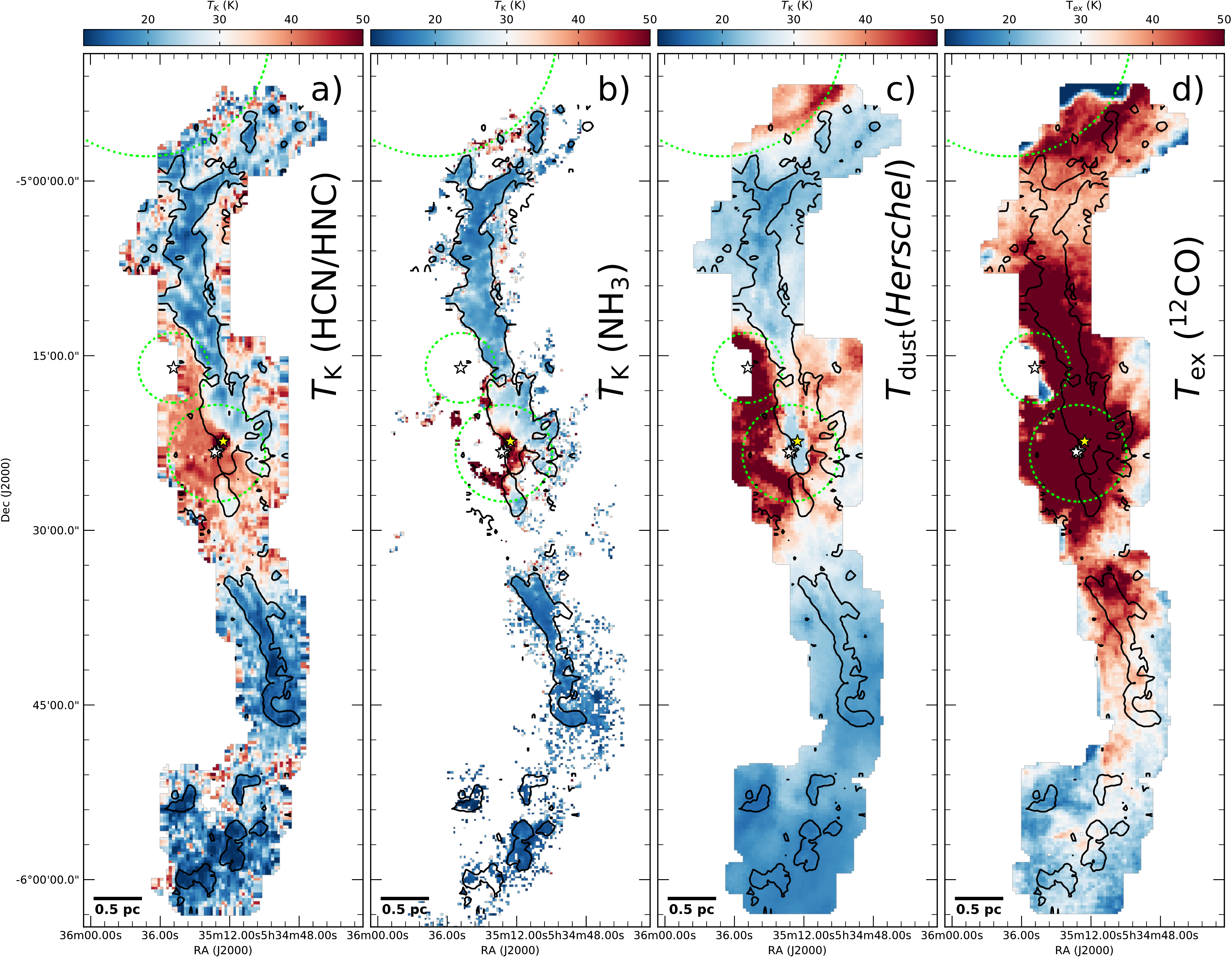}
	\caption{ Comparison of different $T_\mathrm{K}$ estimates along the ISF: {\bf (a)} $T_\mathrm{K}$(HCN/HNC) (this work), {\bf (b)} ammonia-derived gas kinetic temperatures $T_\mathrm{K}$(NH$_3$) \citep{FRI17}, {\bf (c)} Herschel dust temperatures  $T_\mathrm{dust}$ \citep{LOM14}, and {\bf (d)} $^{12}$CO (1--0) excitation temperatures $T_{ex}(^{12}\mathrm{CO})$ \citep{NAK12,SHI14}, all represented with the same colour scale. 
		We note that both dust temperature and column density are underestimated towards the OMC-1 and BN/KL regions as a result of saturation effects and the lack of proper SED fits in the Herschel maps provided by \cite{LOM14}.
		We also note the rapid decrease in $T_{ex}(\mathrm{CO})$ at the northern end in our maps, which is likely due to subthermal excitation of the $^{12}$CO (i.e., $T_\mathrm{ex}(^{12}\mathrm{CO})<T_\mathrm{K}$) lines at the low densities expected in this region.
		To facilitate the comparison of these figure, we indicate the intensity contour with $I$(N$_2$H$^+$)=1.5~K~km~s$^{-1}$ in panels (b-d). Circles and stars are similar to Fig.~\ref{fig:ISF_maps}.
	}
	\label{fig:otherTkmaps}
\end{figure*}

	The temperature structure of the gas and dust in the Orion A cloud have been widely investigated in the past using large-scale molecular line and continuum observations \citep{NAK12,SHI15,NIS15,FRI17}. 
	Figure~\ref{fig:otherTkmaps} shows our $T_\mathrm{K}$(HCN/HNC) measurements (Fig.~\ref{fig:otherTkmaps}) together with other classical temperature estimates based on observations of NH$_3$ (Fig.~\ref{fig:otherTkmaps}b), $^{12}$CO (Fig.~\ref{fig:otherTkmaps}c), and Herschel (Fig.~\ref{fig:otherTkmaps}d)
	within the same area as our IRAM data.
	In Sections~\ref{sec:HCNvsNH3}-\ref{sec:HCNvsCO} we quantify the performance and dynamic range of our empirical $T_\mathrm{K}$(HCN/HNC) measurements as proxy of the gas kinetic temperatures in the ISM in comparison to these previous results.

\subsubsection{HCN/HNC versus NH$_3$}\label{sec:HCNvsNH3}

\begin{figure*}[ht!]
	\centering
	\includegraphics[width=\linewidth]{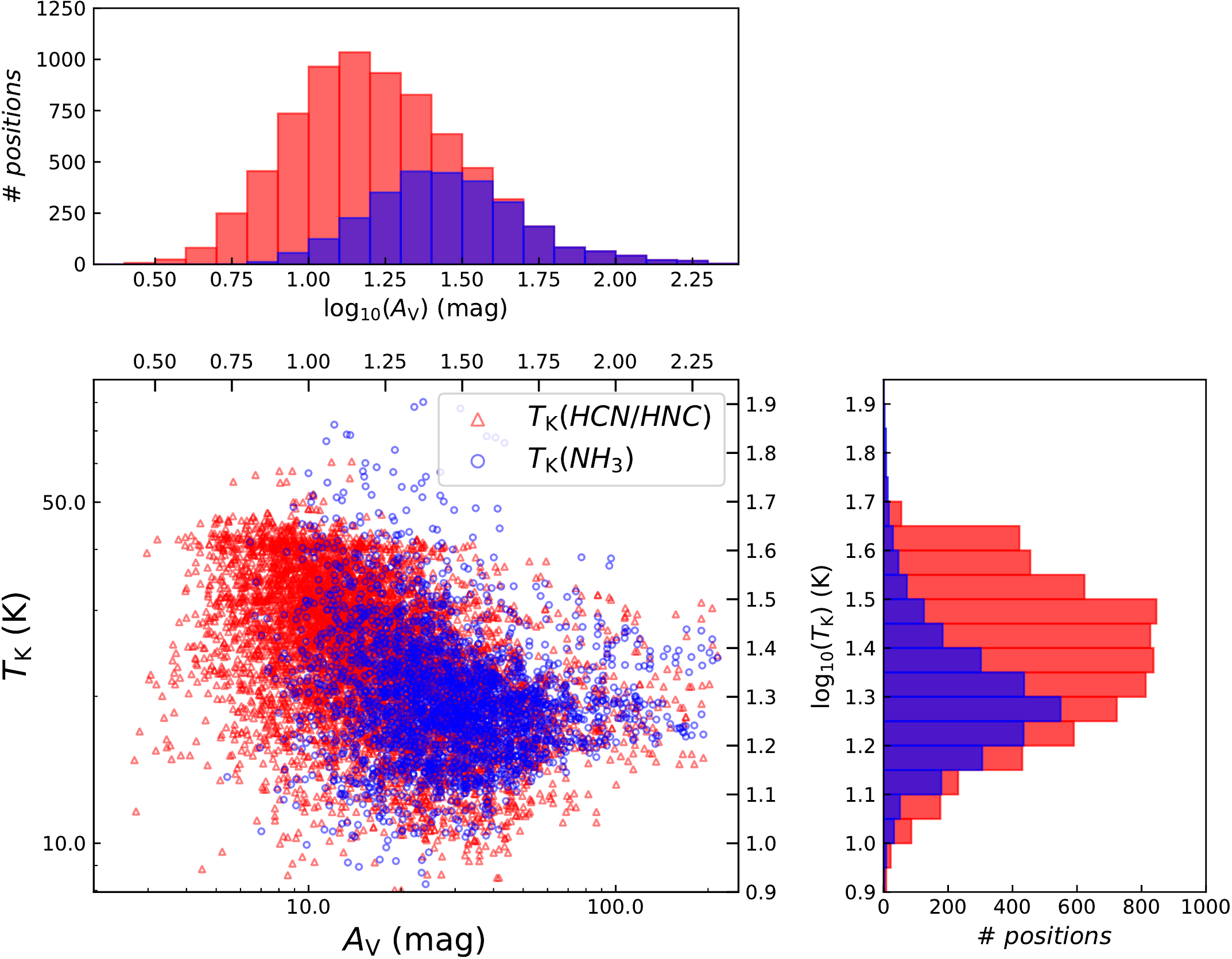}
	\caption{{\bf (Central panel)} Gas kinetic temperatures
		derived using both $T_\mathrm{K}$(HCN/HNC) (red triangles) and $T_\mathrm{K}$(NH$_3$) estimates as function of total gas column density \citep{LOM14} in the ISF region. For simplicity, we display only positions outside the Orion HII nebula  (i.e., $I$(H41$\alpha$)~$\le$~1.0~K~km~s$^{-1}$).
		{\bf (Lateral panels)} Histograms for the distribution of the temperatures (right subpanel) and column densities (upper subpanel) traced by each of these $T_\mathrm{K}$(HCN/HNC) (red bars) and $T_\mathrm{K}$(NH$_3$) (blue bars) measurements. Note that the histograms are binned in log-space.
	}
	\label{fig:ISF_histo}
\end{figure*}

 Together with their corresponding maps (Figs.~\ref{fig:otherTkmaps}a-b),
in Figure~\ref{fig:ISF_histo}~(central panel) we illustrate the temperature variations in the predicted $T_\mathrm{K}$(HCN/HNC) values in our IRAM30m observations with respect to the measured total gas column densities (in units of A$_V$) derived by \citet{LOM14} in Orion using {\it Herschel} observations (red triangles)\footnote{The originally derived A$_K$ values provided by \citet{LOM14} were converted into A$_V$ units following the same procedure described in \citet{HAC17b}.}. For simplicity we display only those positions outside the Orion HII nebula  that are identified by presenting $I$(H41$\alpha$)~$\le$~1.0~K~km~s$^{-1}$. We observe a systematic decrease in recovered $T_\mathrm{K}$(HCN/HNC) gas temperatures at increasing column densities, from about $T_\mathrm{K} \sim $~35~K at A$_V\sim 5$~mag to $T_\mathrm{K} \sim $~12~K at A$_V\sim 50$~mag, as expected for an externally irradiated cloud. At higher column densities, our estimates appear to be also sensitive to the internal gas heating that is produced by the embedded protostars within clouds such as OMC-2/3 (see the warm spots in our maps in these regions).

Maps (Fig.~\ref{fig:otherTkmaps}b) and statistics (Fig.~\ref{fig:ISF_histo}) both highlight the enhanced dynamic range of our $T_\mathrm{K}$(HCN/HNC) temperature estimates.
From the total 8650 Nyquist spectra in our observations, 8092 of these positions (93\%) present  $I$(HCN)~,~$I$(HNC)~$\ge$~1.0~K~km~s$^{-1}$ as suitable for our temperature calculations. These numbers contrast with the much more limited  NH$_3$ measurements, only present in 3469 positions within the same region, or 40\% of the total map coverage. In the histograms shown in Fig.~\ref{fig:ISF_histo} we display a direct comparison of the temperature (right panel) and column density (upper panel) regimes recovered by these $T_\mathrm{K}$(HCN/HNC) (red) and $T_\mathrm{K}$(NH$_3$) (blue) measurements. Similar to NH$_3$, our HCN/HNC observations homogeneously trace the low gas temperatures ($T_\mathrm{K} \lesssim $~20~K) of high column density (A$_V > 20$~mag) material. In a clear improvement compared to NH$_3$, however, our $T_\mathrm{K}$(HCN/HNC) temperature measurements show a significantly higher sensitivity towards warmer regions ($T_\mathrm{K} > $~30~K) at low column densities A$_V < 10$~mag.

\subsubsection{HCN/HNC versus dust}\label{sec:HCNvsdust}

The use of dust effective temperatures $T_\mathrm{dust}$ obtained by different {\it Herschel} observations \citep{LOM14} allows the study of the gas kinetic temperatures at different density regimes in clouds not detected in NH$_3$.
	This comparison can potentially be altered by the density dependence of the gas-to-dust thermal coupling \citep[e.g.][]{GOL01} as well as by the distinct line-of-sight effects weighted by these two measurements.
Despite these caveats, the relative variations of these two observables could be used to quantify the performance of this technique in a wide range of temperatures and densities, at least in a first-order approximation. 

We compared gas and dust temperature measurements from Figures \ref{fig:otherTkmaps}a and  \ref{fig:otherTkmaps}c, respectively.
Overall, $T_\mathrm{K}$(HCN/HNC) and $T_\mathrm{dust}$ estimates show roughly similar variations in our maps. 
Significant differences between these estimates (and NH$_3$) are seen under extreme conditions, both  at low temperatures T$_\mathrm{K}<$~15~K in some of the densest regions within this cloud (denoted by their N$_2$H$^+$ emission) and above T$_\mathrm{K}>$~50~K in the warmest nebulas in this cloud (e.g. the ONC).
Nonetheless, we observe a good correspondence between our $T_\mathrm{K}$(HCN/HNC) estimates and the derived $T_\mathrm{dust}$ values throughout the ISF, particularly at intermediate and low column densities.

\begin{figure}
	\centering
	\includegraphics[width=0.9\linewidth]{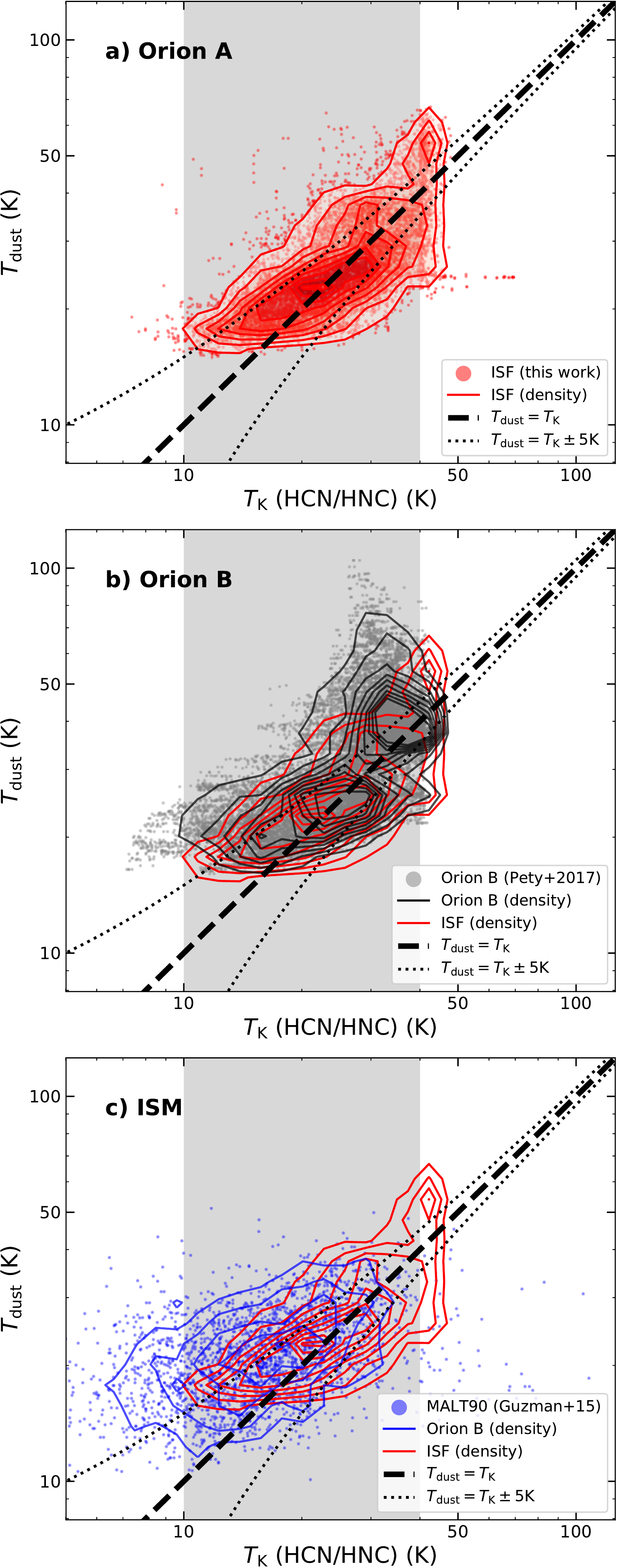}
	\caption{Comparison between the $T_\mathrm{K}$(HCN/HNC) gas kinetic temperatures predicted by our empirical method and the observed dust temperatures in different environments and surveys: {\bf (a)} Orion ISF (this work), {\bf (b)} Orion B \citep{PET16}, and {\bf (c)} MALT90 sample \citep{FOS11,FOS13,JAC13,GUZ15}. We indicate the direct $T_\mathrm{K}=T_\mathrm{dust}$ (dashed line) and $T_\mathrm{K}=T_\mathrm{dust}\pm5$ (dotted lines) correlations in all panels. The optimal temperature range for the application of our method, between $\sim$[10,40]~K, is highlighted in grey in all panels.
	}
	\label{fig:ISM_HCNvsHNC}
\end{figure}

In Figure~\ref{fig:ISM_HCNvsHNC}a, we display a point-by-point comparison between the gas and dust temperature estimates at all positions with $I$(HCN)~,~$I$(HNC) $\ge$~1.0~K~km~s$^{-1}$ in the ISF outside of the ONC nebula. We find an good correlation between the $T_\mathrm{K}$(HCN/HNC) and $T_\mathrm{dust}$ variations in the ISF both in relative and absolute terms, almost following a 1:1 relationship (dashed line) with no significant dependence on the corresponding total gas column density. More than 50\% of our $T_\mathrm{K}$(HCN/HNC) estimates present values that are similar to their corresponding $T_\mathrm{dust}$ measurements with deviations of less than $\pm$~5~K (see dotted lines). This agreement is particularly striking within the optimal temperature range of our method.  

Taken together, theoretical (Sect.~\ref{sec:Tk_calibration}) and empirical (Fig.~\ref{fig:ISF_histo}) arguments demonstrate the robustness of the observed total  $I\mathrm{(HCN)}$-to-$I\mathrm{(HNC)}$ integrated intensity ratio as direct measurement of the $T_\mathrm{K}$ gas kinetic temperatures in a wide range of temperature and column density regimes. 
Only limited by the extension of our maps, this simple technique benefits from the bright emission of HCN and HNC isomers in dense and diffuse media \citep{PET16,KAU17}. Easily obtained in regular millimeter line observations, 
this simple observable opens a new window on the study of the gas thermal properties of clouds at parsec scales.

\subsubsection{HCN/HNC versus CO}\label{sec:HCNvsCO}
	
	\begin{figure}
		\centering
		\includegraphics[width=1.0\linewidth]{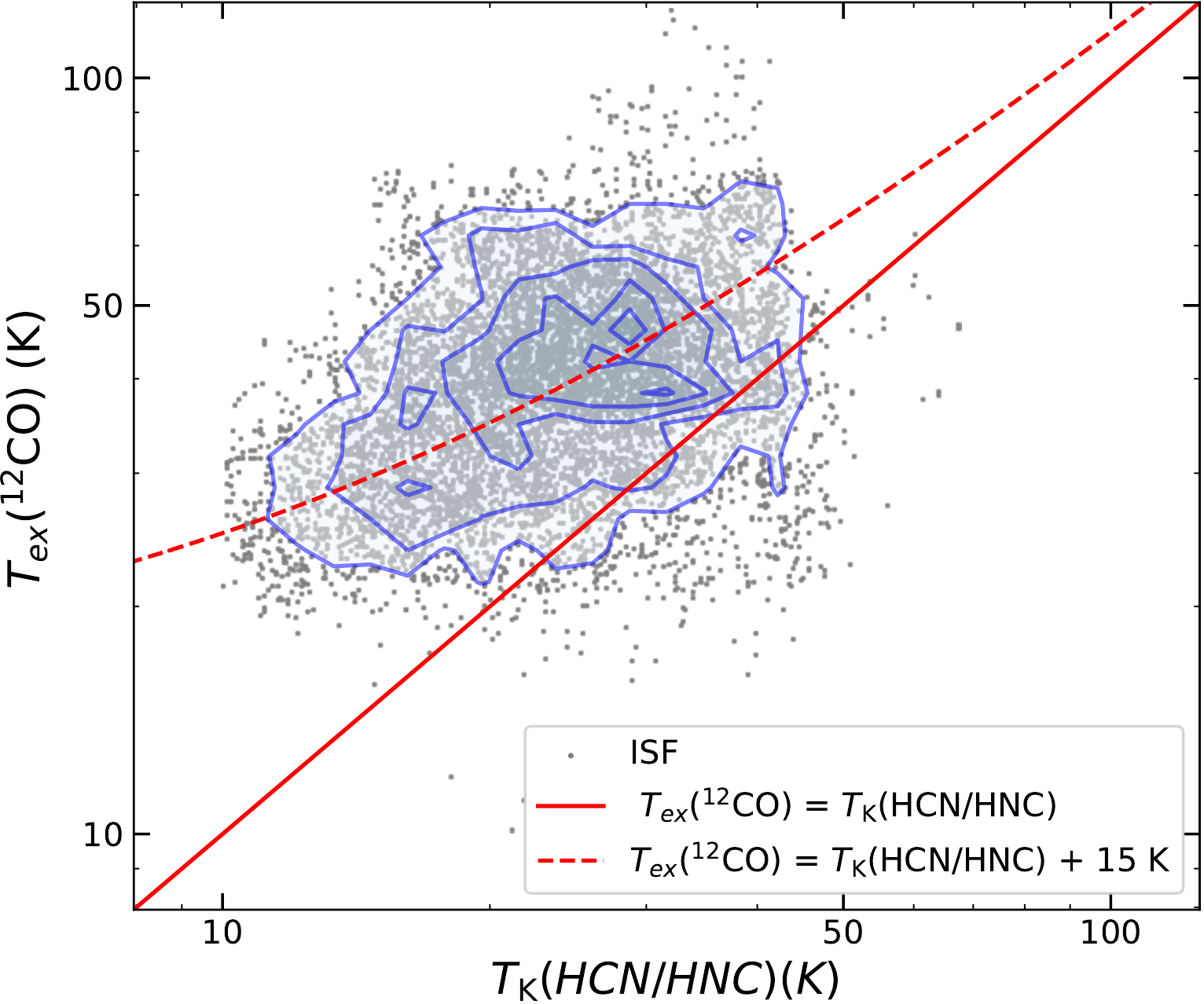}
		\caption{
			Comparison between our new $T_\mathrm{K}$(HCN/HNC) temperature estimates and the  $^{12}$CO (1--0) excitation temperatures $T_{ex}(^{12}\mathrm{CO})$ \citep{NAK12,SHI14} in all positions throughout the ISF outside the ONC nebula Orion HII nebula (i.e., $I$(H41$\alpha$)~$\le$~1.0~K~km~s$^{-1}$; grey dots).
			The solid line indicates the linear correlation between these two temperatures estimates  $T_{ex}(^{12}\mathrm{CO})$~=~$T_\mathrm{K}$(HCN/HNC).
			Although they are positively correlated (see density contours in blue), we note that the $T_{ex}(^{12}\mathrm{CO})$ temperature estimates are systematically higher than the corresponding $T_\mathrm{K}$(HCN/HNC) values at the same position. For illustrative purposes we display the $T_{ex}(^{12}\mathrm{CO})\sim T_\mathrm{K}$(HCN/HNC)+15~K line denoted by a dashed red line in our plot.
		}
		\label{fig:ISM_HCNvsCO}
	\end{figure}
	
	As for the dust, observations of diffuse tracers such as $^{12}$CO provide information of the thermal structure of clouds at large scales.
	For any emission line, its observed peak temperature ($T_\mathrm{peak}$) is connected to its excitation temperature ($T_\mathrm{ex}$) and to its line opacity ($\tau$), according to the radiative transfer equation $T_\mathrm{peak}=\left( J(T_\mathrm{ex})-J(T_\mathrm{bg}) \right)(1-e^{-\tau})$, where $J(T)=\frac{h\nu/k}{exp(h\nu/kT)-1}$ and $T_\mathrm{bg}$ corresponds to the background temperature. In the optically thick regime ($\tau\gg 1$), this equation can be simplified to obtain a direct measurement of the gas excitation temperature as:
	\begin{equation}
	T_\mathrm{ex}=\frac{h\nu/k}{ln\left(\frac{h\nu}{k(T_\mathrm{peak}+J(T_\mathrm{bg}))} \right)+1}\label{eq:Tex}
	\end{equation}
	Favoured by its typically high opacities in dark clouds ($\tau(^{12}\mathrm{CO})> 10$), the $^{12}$CO (J=1--0) transition ($\nu=115.271$~GHz) is one of the most commonly used tracers of the gas temperatures in the ISM under the assumption of LTE conditions for which $T_\mathrm{K} = T_\mathrm{ex}(^{12}\mathrm{CO})= f(T_\mathrm{mb}(^{12}\mathrm{CO}))$.
	
	Many studies have explored the excitation conditions of $^{12}$CO using large-scale observations in Orion \citep[e.g.][ among others]{NAK12,SHI14,NIS15}.
	Figure~\ref{fig:ISF_Tkmaps}~d shows the excitation temperature $T_\mathrm{ex}(^{12}\mathrm{CO})$ of the $^{12}$CO (J=1--0) line according to Eq.\ref{eq:Tex} using the line peak temperatures provided by \citet{SHI14} within the same area as was explored by our IRAM30m observations. As necessary (although perhaps not sufficient) condition, the high gas densities expected throughout the ISF favours the thermalization of the observed $^{12}$CO lines in most of the observed positions within our maps.
	A quick comparison with Figs.\ref{fig:ISF_Tkmaps}~a-c reveals large differences between these CO temperature estimates and the equivalent HCN, NH$_3$, and dust measurements, both in terms of distribution and absolute values. Overall, $T_\mathrm{ex}(^{12}\mathrm{CO})$ presents warmer temperatures throughout the entire ISF. In particular, the observed $T_\mathrm{ex}(^{12}\mathrm{CO})$ temperatures are dominated by the strong heating effects around the ONC, M43, and NGC1977 nebulae \citep[see dotted green lines; see][]{NIS15}. In regions such as OMC-1 or OMC-2 these $T_\mathrm{ex}(^{12}\mathrm{CO})$ estimates also show a small dynamic range in temperature and are less sensitive to the cloud structure showing almost no variation compared to the distribution of denser and colder material traced by other gas or dust measurements.
	
	Figure~\ref{fig:ISM_HCNvsCO} quantifies the differences between the derived $T_\mathrm{K}\mathrm{(HCN/HNC)}$ and $T_\mathrm{ex}(^{12}\mathrm{CO})$ temperatures in all positions sampled in our ISF observations outside the ONC (see Fig.~\ref{fig:otherTkmaps}). 
	As illustrated by this plot, $T_\mathrm{ex}(^{12}\mathrm{CO})$ presents systematically higher values than $T_\mathrm{K}\mathrm{(HCN/HNC)}$, typically warmer by about 15~K (see dashed line in the plot).
	Despite a large scatter, $T_\mathrm{K}\mathrm{(HCN/HNC)}$ and $T_\mathrm{ex}(^{12}\mathrm{CO})$ temperatures show a positive correlation within the $T_\mathrm{K}\mathrm{(HCN/HNC)}\sim$~10-50~K regime. Although different in absolute terms, this 
	parallel evolution suggests a physical link between these temperature estimates within our maps.
	
	The observed differences between these $T_\mathrm{K}\mathrm{(HCN/HNC)}$ and $T_\mathrm{ex}(^{12}\mathrm{CO})$ temperature estimates (Fig.~\ref{fig:ISM_HCNvsCO}) can be explained by the different cloud depths that are traced in each of these measurements. By construction, the reported $T_\mathrm{ex}(^{12}\mathrm{CO})$ values estimate the gas temperatures at the cloud depth in which the $^{12}$CO line becomes optically thick (i.e., $\tau(^{12}\mathrm{CO})> 1$). As a results of the large $^{12}$CO abundance, the $^{12}$CO (1--0) line quickly saturates at low column densities in the outermost warmer layers of dense regions like Orion. On the other hand, optically thinner transitions such as HCN (1--0) and HNC (1--0) provide information from deeper and therefore colder layers of gas within these regions (e.g. see anti-correlation between $T_\mathrm{K}$ and $A_\mathrm{V}$ in Fig.~\ref{fig:ISF_histo}, central panel).
	Sensitive to distinct parcels of gas the simultaneous study of  $T_\mathrm{K}\mathrm{(HCN/HNC)}$ and $T_\mathrm{ex}(^{12}\mathrm{CO})$ could be used to investigate different temperature layers and regimes within clouds.

\subsection{Universality: HCN/HNC as a probe of the gas kinetic temperatures in the ISM}\label{sec:universal}

In addition to studying Orion, it is fundamental to explore whether this newly proposed thermometer can be extrapolated for the analysis of other star-forming regions.
Similar to the ISF (Sect.~\ref{sec:application}), we investigate the correlation between the $T_\mathrm{K}$(HCN/HNC) and $T_\mathrm{dust}$ temperatures in the Orion B cloud in Figure~\ref{fig:ISM_HCNvsHNC}b. To produce this plot, we resampled and combined the HCN/HNC intensities provided by \citet{PET16} with the corresponding dust effective temperatures derived in \citet{LOM14} (grey dots). 
These molecular data expand over a total area of $\sim$~1~deg$^2$ within the central region of the Orion B cloud, covering a wide range of star-forming and thermal conditions \citep[see][]{PET16}. We converted each HCN/HNC intensity ratio into its corresponding gas kinetic temperature following our empirical prescriptions.
Altogether, we find a good correspondence between the $T_\mathrm{K}$(HCN/HNC) values in Orion B predicted by our method and the expected $T_\mathrm{dust}$ variations (black contours). Deviations from this correlation are observed at temperatures $>$~30~K, likely due to the limitations of our method in combination of line-of-sight effects.  
Nonetheless, we find a direct correspondence between the temperature estimates in Orion B and our Orion A results (red contours) in dynamic range and absolute variations. 

Testing the robustness of this novel technique in other clouds is 
hindered by the limited availability of simultaneous HCN, HNC, and $T_\mathrm{dust}$ maps.
In the absence of large-scale observations similar to our Orion data, we have compared the predicted gas kinetic temperatures obtained using HCN/HNC line ratios with additional dust temperature estimates extracted from different molecular surveys across the Milky Way.  
In particular, we combined the observed $I\mathrm{(HCN)}/I\mathrm{(HNC)}$  values in dense clumps provided by the MALT90 survey \citep{FOS11,FOS13,JAC13} and $T_\mathrm{dust}$ measurements described by \citet{GUZ15} (3216 sources). 
Several caveats should be considered for this comparison. Each MALT90 target corresponds to a parsec-like size clump that is unresolved within this survey beam, and is sometimes internally heated by embedded sources. Multiple gas and dust temperatures are therefore expected to be convolved within a single measurement. The correlation between the gas and dust estimates could be largely affected by the different sensitivity and weights of these measurements within this beam. As result, this survey does not allow the inspection of the temperature structure of individual clumps. Instead, we use this analysis to explore the $I\mathrm{(HCN)}/I\mathrm{(HNC)}$ intensity variations in different Galactic environments in a statistical manner.

We display the comparison between the $T_\mathrm{K}$(HCN/HNC) and $T_\mathrm{dust}$ temperatures in all the MALT90 clumps with simultaneous HCN, HNC, and dust measurements in Figure~\ref{fig:ISM_HCNvsHNC}c (blue dots). 
Despite their observational caveats, the strong temperature dependence of the HCN/HNC ratio still prevails within the MALT90 sample. Although obviously noisier than our high-resolution data, we observe a global correlation between the $T_\mathrm{dust}$ and $T_\mathrm{K}$(HCN/HNC) measurements (blue contours) following the same trend as defined in our ISF data (red contours), particularly at T$_\mathrm{K}>$~10~K.

The close correspondence between the predicted $T_\mathrm{K}$(HCN/HNC) gas temperatures and the observed $T_\mathrm{dust}$ dust measurements confirms the validity of the HCN/HNC line ratio as proxy of the gas temperatures beyond the ISF. 
The good agreement between these independent observables in regions such as Orion B (Fig.~\ref{fig:ISM_HCNvsHNC}b) highlights the good performance of the proposed $I\mathrm{(HCN)}/I\mathrm{(HNC)}$ values as proxy of the gas kinetic temperatures at cloud scales. Moreover, that a similar correlation is observed in the MALT90 sample allows us to extend these local results to different environments throughout the Milky Way.
These results denote this newly proposed $I\mathrm{(HCN)}$-to-$I\mathrm{(HNC)}$ line ratio as a reliable and robust estimate for the gas kinetic temperatures of the molecular ISM.

\section{Discussion}
\subsection{ HCN versus HNC emission properties as a function of temperature}\label{sec:emission}

\begin{figure}[ht!]
	\centering
	\includegraphics[width=\linewidth]{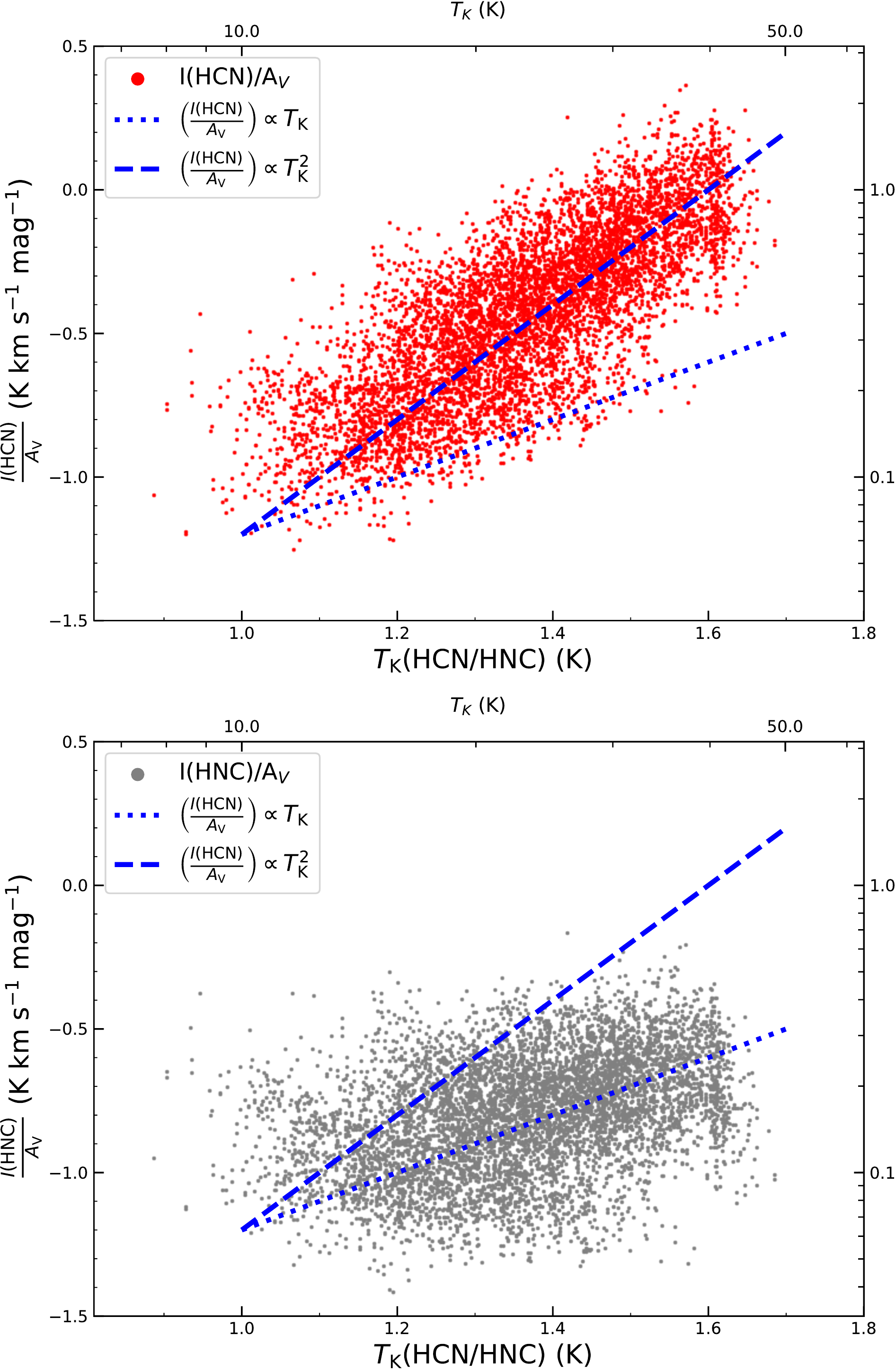}
	\caption{
		Specific HCN {\bf (top)} and HNC {\bf (bottom)} intensities normalized by column density (i.e., $I$(X)/A$_V$) as function of the $T_\mathrm{K}$(HCN/HNC) gas temperatures
	for all positions in our ISF maps with $I$(HCN),~$I$(HNC)~$\ge$~1.0~K~km~s$^{-1}$ outside the ONC, that is, with $I$(H41$\alpha$)~$\le$~1.0~K~km~s$^{-1}$. For comparison, linear (blue dotted line) and quadratic (blue dashed line) variations of the normalized line intensities with temperature are indicated in all  subpanels.  
	}
	\label{fig:Luminosities}
\end{figure}

The advent of broad-band receivers has popularized the use of simultaneous HCN and HNC observations in a broad range of astrophysical studies.
Variations in the observed the HCN/HNC intensity ratio are employed as indicator of the evolutionary stage of high-mass star-forming regions in the Milky Way \citep{JIN15,COL18}.
High HCN/HNC values are also observed in Seyfert \citep{PER07} and starburst galaxies \citep{BEM19}.
Systematic changes of this ratio are also observed between the central and disk regions of nearby galaxies \citep{JIM19}.
At smaller scales, significant changes in the distribution and ratio of these two isomers have been reported from planetary nebulae to \citep{BUB19} to protoplanetary disks \citep{GRA15}. The robustness of results derived from our Orion observations suggest that these intensity variations may be related to changes in the gas kinetic temperatures.

We explore the influence of temperature on the observed HCN and HNC intensities in Figure \ref{fig:Luminosities}.
We display the individual HCN (top panel) and HNC (bottom panel) integrated intensities normalized by the local extinction measurements derived using {\it Herschel} observations \citep{LOM14} as a function of the gas kinetic temperature ($T_\mathrm{K}$(HCN/HNC)) for all positions throughout the ISF outside the ONC. 
The use of this extinction normalization (also known as specific intensity) allows us to isolate the enhanced abundance and temperature effects on the observed line intensities from their similar expected increase as a function of column density. 

Figure~\ref{fig:Luminosities} shows that the specific HCN and HNC intensities systematically increase at higher gas kinetic temperatures $T_\mathrm{K}$. 
However, the comparison of their individual temperature dependence shows clear differences between these two molecular tracers. HNC (bottom panel) grows linearly with temperature ($\propto T_\mathrm{K}$; blue dotted line). Much steeper, HCN (top panel) shows a quadratic dependence with the gas kinetic temperature (i.e., $\propto T_\mathrm{K}^2$; blue dashed line). The strong temperature dependence observed in HCN produces changes of more than an order of magnitude in the intensity of this latter tracer per unit of column density within the range of temperatures we considered here. 

The different behaviours of the observed HCN and HNC emission properties shown in Fig.~\ref{fig:Luminosities} can be explained by the combination of excitation and chemical effects. Both observed HCN and HNC intensities are enhanced by the increasing excitation conditions at higher temperatures (linear dependence).
In addition to this, the observed HCN intensities are boosted (quadratic dependence) by the increasing production and abundance of this isomer in lukewarm conditions above $>$~15~K (see Sect.~\ref{sec:Ebarrier}).

\subsection{Interpretation of HCN observations in extragalactic studies}\label{sec:extragalactic}

\begin{figure}[t!]
	\centering
	\includegraphics[width=\linewidth]{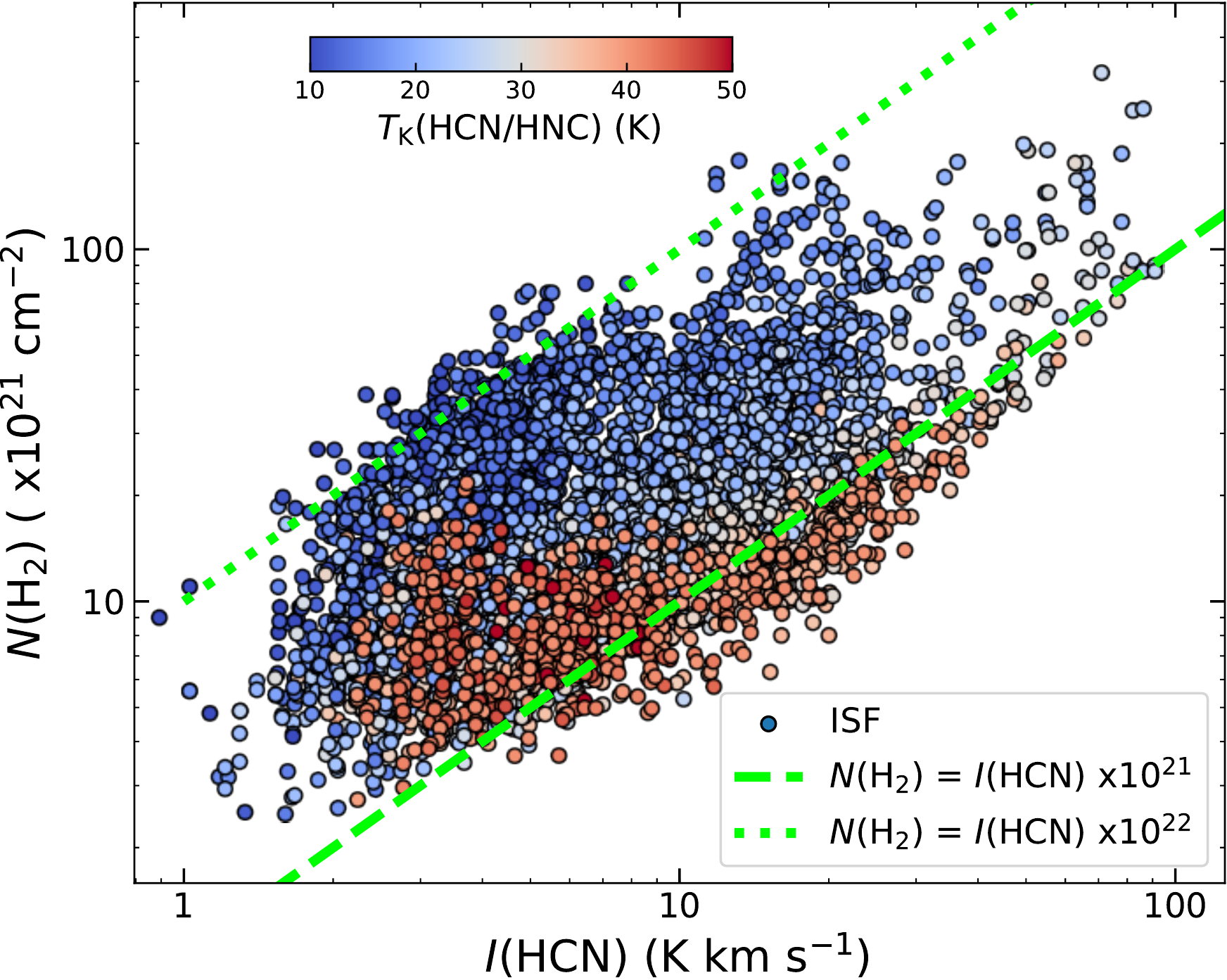}
	\caption{Correlation between the observed total HCN intensities $I$(HCN) and the total gas column density $N(\mathrm{H}_2)$ for all positions in our maps, colour-coded by their gas kinetic temperature (see scale bar in the top left corner).  We note the large spread in our data 
		(up to 1~dex) for a given column density $N(\mathrm{H}_2)$ or intensity $I$(HCN) value. 
	}
	\label{fig:HCN_Xfactor}
\end{figure}

The magnitude of the above temperature variations significantly alters the correlation between the observed HCN intensities $I$(HCN) and total gas column densities $N(\mathrm{H}_2)$, that is, the so-called X-factor for HCN.
Under normal excitation conditions, the HCN intensities are expected to grow with $N(\mathrm{H}_2)$ through the intrinsic correlation between $N(\mathrm{H}_2)$ and $n(\mathrm{H}_2)$ \citep{BIS19}.
However, the resulting intensity at a given column density can strongly depend on the gas temperature (see Sect.\ref{sec:emission}).
We quantify this effect in Figure~\ref{fig:HCN_Xfactor} by comparing these two observables assuming a standard $N(\mathrm{H}_2)/A_\mathrm{V}=0.93\times 10^{21}  $~cm$^{-2}$~mag$^{-1}$ conversion \citep{BOH78}. To illustrate its temperature dependence, each of our observations is colour-coded according to its corresponding $T_\mathrm{K}$(HCN/HNC) value. 
At low gas temperatures, the reported $N(\mathrm{H}_2)$ describes an almost linear relationship with the $I$(HCN) intensities (e.g. $N(\mathrm{H}_2)[\mathrm{cm}^{-2}]\sim I(\mathrm{HCN})[\mathrm{K~km~s}^{-1}]\times 10^{22}$ at 10~K, see dotted line).
However, this correlation is systematically shifted towards higher $I$(HCN) values at increasing gas tempetures (e.g. $N(\mathrm{H}_2)[\mathrm{cm}^{-2}]\sim I(\mathrm{HCN})[\mathrm{K~km~s}^{-1}]\times 10^{21}$ at 40~K, see dashed line). As a result, each $I$(HCN) intensity value represents range of column densities greater more than a factor of 5. Similarly, a fixed $N(\mathrm{H}_2)$ column shows a variation of almost a 1~dex in intensity depending on the intrinsic gas temperature (see colour-coded temperatures in this figure).

\begin{figure}[ht!]
	\centering
	\includegraphics[width=\linewidth]{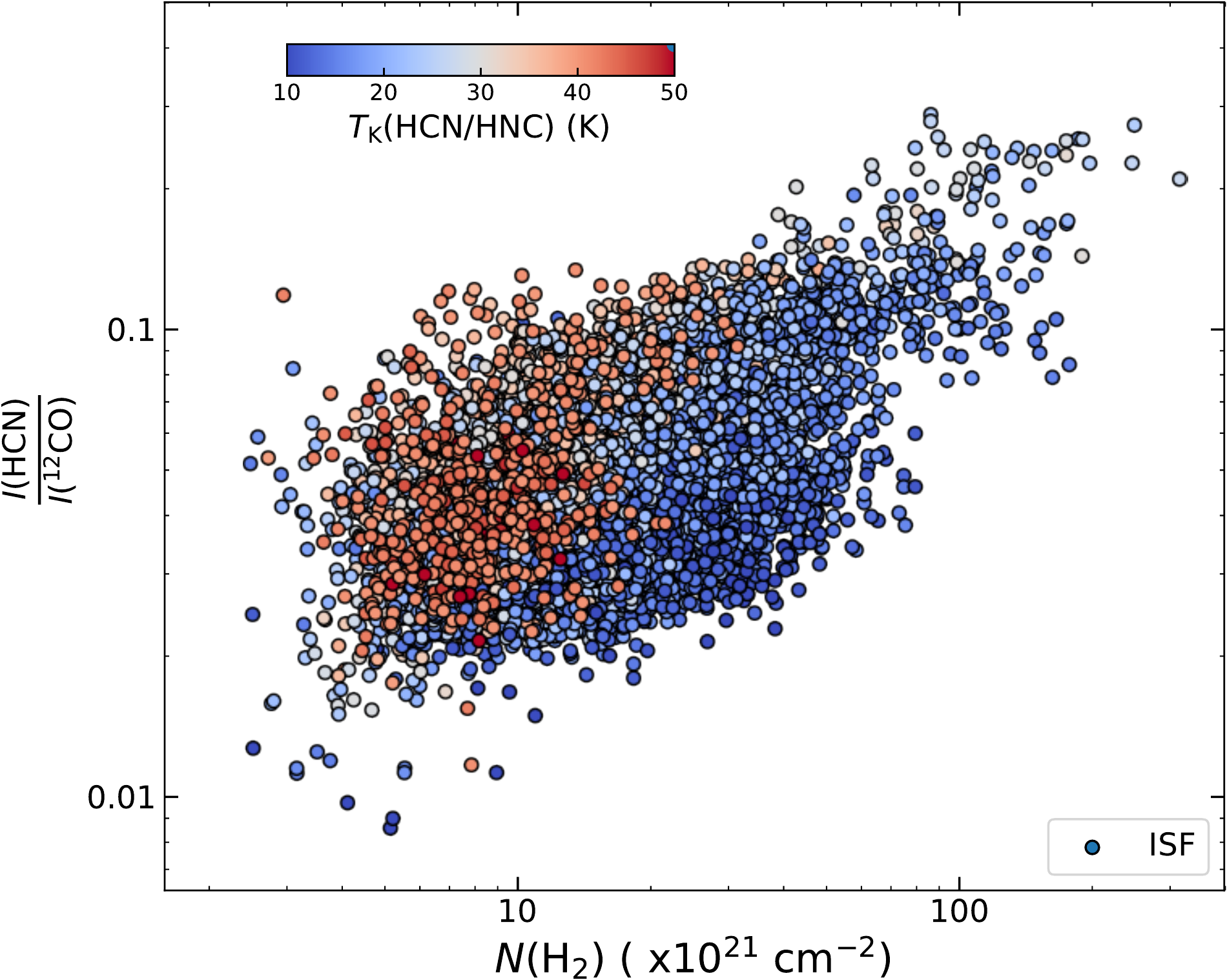}
	\caption{
		Total gas column density $N(\mathrm{H}_2)$ as a function of the HCN (1--0) vs $^{12}$CO (1--0) line intensity ratio ($I\mathrm{(HCN)}/I(^{12}\mathrm{CO})$)  for all positions in our ISF maps outside the ONC with $I$(H41$\alpha$)~$\le$~1.0~K~km~s$^{-1}$. All points are colour-coded by the corresponding $T_\mathrm{K}$(HCN/HNC) value (see the scale bar in the top left corner).
	}
	\label{fig:HCNvsCO_H2}
\end{figure}

	Although less prominent, the reported temperature enhancements of the HCN abundances (Fig.~\ref{fig:AbsAbun}) and intensities (Fig.~\ref{fig:Luminosities}) also change the line emission ratio of this molecule with respect to other tracers such as CO. In Figure~\ref{fig:HCNvsCO_H2} we show the HCN (1--0) (this work) and $^{12}$CO (1--0) \citep{NAK12,SHI14} line ratio $I\mathrm{(HCN)}/I(^{12}\mathrm{CO})$ as a function of total gas column density $N(\mathrm{H}_2))$ within our maps (see Sect.~\ref{sec:HCNvsCO} for further details). As expected, we observe a gradual increase in $I\mathrm{(HCN)}/I(^{12}\mathrm{CO})$ with $N(\mathrm{H}_2)$. However, we identify a systematic spread in both axes that is associated with temperature effects (see coloured points). For the majority of the positions sampled in this plot, the same $I\mathrm{(HCN)}/I(^{12}\mathrm{CO})$ line ratio describes a wide range of column densities and therefore gas masses. Similarly, a single $N(\mathrm{H}_2)$ value can present a variation up to a factor of $\sim 5$ in its corresponding $I\mathrm{(HCN)}/I(^{12}\mathrm{CO})$ line ratio that is largely associated to the effects of temperature on the emission properties of the HCN lines.

Our results challenge the use of HCN observations in extragalactic sources. The pioneering works by \citet{GAO04} suggested the observed HCN luminosities
 as (linear) indicator of the total star formation rate from nearby spiral galaxies to the most distant luminous (LIGs) and ultra luminous infrared galaxies (ULIGs) \citep[see also][]{WU05} \footnote{In extragalactic observations, the integrated emission of a given molecule is typical described in terms of luminosities.	Line intensities ($I$) and luminosities ($L$) are related by $L=I\times A$, where $A$ usually defines the solid angle subtended by the solid source area \citep[e.g. see Eqs.1-4 in ][]{GAO04_2}. Although indeed different when individual measurements are considered, intensity and luminosity ratios become equivalent terms, that is, $\frac{I\mathrm{(HCN)}}{I\mathrm{(HNC)}}=\frac{L\mathrm{(HCN)}}{L\mathrm{(HNC)}}$, in observations of a single source and when a similar beamsize is used, such as in our ISF maps.}. This interpretation assumes the HCN  intensities as direct measurement of the total dense ($n$(H$_2$)$> 10^4$~cm$^{-3}$ at A$_V > 8^{mag}$) and star-forming gas in these extragalactic environments \citep[i.e., M$_{dense}\propto $~$I$(HCN); see][for a full discussion]{LAD12}. 
Conversely, local observations of resolved molecular clouds demonstrate that the integrated HCN (1--0) intensity at parsec scales is dominated by the contribution of more diffuse gas ($n$(H$_2$)~$< 10^3$~cm$^{-3}$) at low column densities (A$_V < 6^{mag}$) \citep{KAU17}.
Recent calculations suggest that part of these enhanced HCN intensities at low extinctions could be produced by electron collisions \citep{GOL17}.
The strong temperature dependence of the HCN  total integrated intensities reported in our work adds another layer of complexity to this interpretation.
As demonstrated by Fig.~\ref{fig:Luminosities}, the specific HCN intensity per-unit-of-column density (i.e. $\frac{I(\mathrm{HCN})}{A_V}$) produced at temperatures of $\sim$~40~K exceeds the  intensity of the same amount of gas at temperatures of $\sim$~10~K by more than an order of magnitude.
	Moreover, these same temperature effects also increase the HCN emission with respect to more diffuse tracers such as CO as illustrated on Fig.~\ref{fig:HCNvsCO_H2}.
Based on these results we speculate that a non-negligible part of the HCN luminosities in these spiral, LIG, and ULIG galaxies could originate from the increase of the average gas temperature in these extragalactic sources. 
Rather than measuring the content of dense gas converted into stars, the reported HCN luminosities would then reflect the increasing impact of stellar feedback onto the surrounding molecular material in high-mass star-forming regions such as Orion.
Systematic local and extragalactic observations of HCN and HNC tracers are needed to confirm this hypothesis beyond our data.

\section{Conclusions}\label{sec:conclusions}

We characterized the distribution and emission properties of HCN and HNC isomers throughout the Integral Shape Filament (ISF) in Orion using a new set of large-scale HCN and HNC (1-0) IRAM30m maps.
We compared our new IRAM30m data with ancillary GBT-NH$_3$ \citep{FRI17} and {\it Herschel} \citep{LOM14} observations and analysed them with a new set of chemical models. This unique combination allowed us to explore the use of the observed $I$(HCN)/$I$(HNC) intensity ratios as a new chemical gas thermometer for the molecular ISM.

The main results of this work are summarized below:
\begin{enumerate}

\item We find a strong and systematic correlation between the observed $I$(HCN)/$I$(HNC) intensity ratio and the ammonia-derived gas kinetic temperatures $T_\mathrm{K}$ in Orion, showing increasing HCN/HNC values at higher gas temperatures (Sect.~\ref{sec:ratios}).

\item We describe the observed variations of the $I$(HCN)/$I$(HNC) intensity ratios reported in the ISF using a two-part linear function at different high (T$_K\gtrsim$~40~K) and low (T$_K<$~40~K) temperatures (see Sect.~\ref{sec:Tk_calibration}). 
Easily obtained in standard millimeter-line observations, this empirical parametrization allows the use of this $I$(HCN)/$I$(HNC) ratio as a direct proxy of the gas kinetic temperatures $T_\mathrm{K}$(HCN/HNC).

\item Our chemical models demonstrate that the observed $I$(HCN)/$I$(HNC) intensity variations are driven by similar changes in the abundance ratio X(HCN)/X(HNC) of these two isomers due the efficient destruction of HNC via (1) HNC~+~H~$\rightarrow$~HCN~+~H and (2) HNC~+~O~$\rightarrow$~NH~+~CO reactions operating under low-energy barriers $\Delta E_i$. Using observational estimates and model calculations we derived values of $\Delta E_1=$~200~K \citep{GRA14} and $\Delta E_2\sim$~20~K (this work) for reactions (1) and (2), respectively. Our $\Delta E_i$ estimates reduce the energy barriers predicted by classical ab initio calculations by approximately an order of magnitude  (see Sect.~\ref{sec:Ebarrier}).

\item We used our empirical calibration of the observed $I$(HCN)/$I$(HNC) intensity ratio to obtain a large-scale temperature map $T_\mathrm{K}$(HCN/HNC) of the entire Orion ISF (Sect.~\ref{sec:application}). 
Our temperature estimates show consistent results ($<$~5~K) with independent measurements of the dust effective temperature $T_\mathrm{dust}$ derived using {\it Herschel} observations \citep{LOM14}. Interestingly, the bright emission of the HCN and HNC isomers in our maps allows us to obtain systematic temperature measurements within a large dynamic range of column densities ($5\,\mathrm{mag}< \mathrm{A}_V \sim 100\,\mathrm{mag}$) and temperatures (15~K~$\lesssim$~T$_K \lesssim$~40~K). Thus, our results extend previous NH$_3$ measurements that were typically restricted to A$_V>10$~mag and provide  a more complete description of the thermal structure of the diffuse material in clouds like Orion down to A$_V$ of a few magnitudes.

\item In addition to Orion A, we find an excellent correlation between our $T_\mathrm{K}$(HCN/HNC) gas temperature estimates and additional {\it Herschel} $T_\mathrm{dust}$ dust temperature measurements in similar local \citep[Orion B;][]{PET16} and galactic \citep[MALT90;][]{JAC13} surveys  (Sect.~\ref{sec:universal}).
The close correspondence between these two observables in different environments confirms the potential of the $I$(HCN)/$I$(HNC) intensity ratio as a universal thermometer for the molecular ISM.

\item Comparisons with both gas and dust temperature estimates show that the specific intensity of HCN per unit of column density  is rapidly enhanced with temperature (i.e., $I$(HCN)/A$_V\propto$~T$^2$) in comparison to the HNC (e.g. $I$(HNC)/A$_V\propto$~T) due the combined effect of excitation plus chemical reactions (1 and 2) between these two isomers (Sect.~\ref{sec:emission}). These systematic variations of the HCN intensities might affect the interpretation of this molecule as a tracer of the dense gas content in extragalactic observations (Sect.~\ref{sec:extragalactic}). 

\end{enumerate}

\begin{acknowledgements}
	A.H. thanks the insightful discussions with A. Usero and M. Tafalla.
	This work is part of the research programme VENI with project number 639.041.644, which is (partly) financed by the Netherlands Organisation for Scientific Research (NWO). AH thanks the Spanish MINECO for support under grant AYA2016-79006-P.
	  Based on observations carried out with the IRAM30m Telescope. IRAM is supported by INSU/CNRS (France), MPG (Germany) and IGN (Spain).
      This research made use of APLpy, an open-source plotting package for Python \citep{aplpy}.
      This research made use of the WCS tools \citep{WCSpub}.
      This research made use of Astropy, a community-developed core Python package for Astronomy \citep{Astropy}.
      This research made use of TOPCAT \citep{TOPCAT}.
\end{acknowledgements}

%
%

\begin{appendix} 
	
\section{Abundance versus intensity ratios}\label{sec:appendix2}

\begin{figure*}
	\centering
	\includegraphics[width=\linewidth]{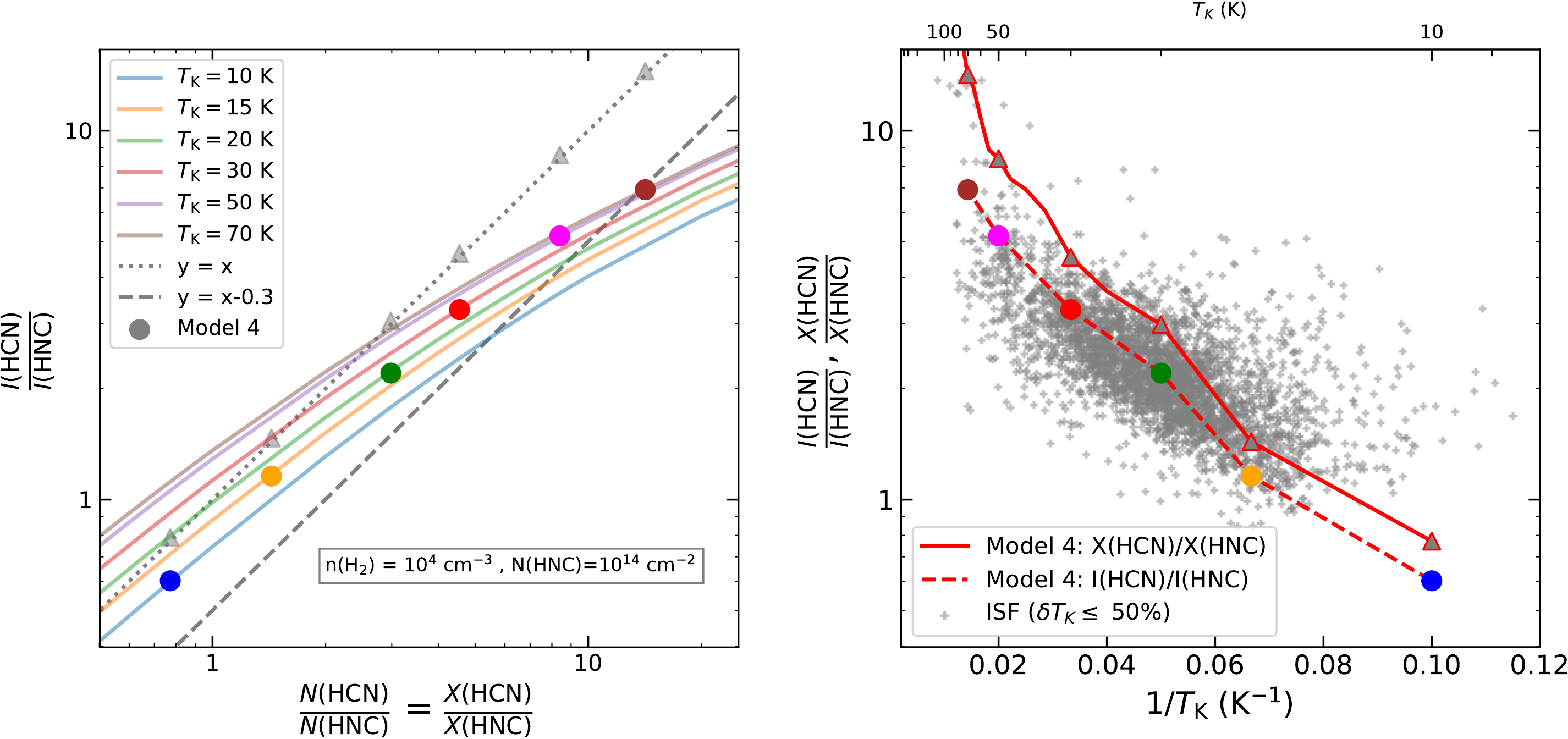}
	\caption{Comparison between the HCN vs HNC intensity ratio ($L(\mathrm{HCN})/L(\mathrm{HNC})$) as function of the abundance ratio ($X(\mathrm{HCN})/X(\mathrm{HNC})$) for different gas kinetic temperatures ($T_\mathrm{K}$). 
	{\bf (Left)} Luminosity variations for  $N(\mathrm{HCN})/N(\mathrm{HNC})=X(\mathrm{HCN})/X(\mathrm{HNC})=[0.1,100]$ for gas temperatures between 10 and 70~K (coloured lines) calculated using RADEX assuming $n(\mathrm{H}_2)=10^4$~cm$^{-3}$ and $N(\mathrm{HNC})=10^{14}$~cm$^{-2}$ (see text). Linear variations are indicated by dotted (x=y) and dashed (y=x-0.3~dex) lines.
The abundance and intensity ratios expected for our model 4 are indicated by solid circles coloured according to their corresponding temperature.
	{\bf (Right)} Differences between the predicted $X(\mathrm{HCN})/X(\mathrm{HNC})$ (solid line) and $L(\mathrm{HCN})/L(\mathrm{HNC})$ (dashed line) values for model 4  as a function of 1/$T_\mathrm{K}$ (solid circles; similar to the left panel) in comparison to our Orion observations (grey crosses), similar to Fig.~\ref{fig:AbsAbun}. We note the close correspondence between these $X(\mathrm{HCN})/X(\mathrm{HNC})$ and $L(\mathrm{HCN})/L(\mathrm{HNC})$ measurements under the physical conditions considered in our models.
}
	\label{fig:LvsX}
\end{figure*}

To characterize the HCN and HNC variations, 
our analysis assumed a direct correspondence between the predicted abundance ratios of these species (i.e. $X(\mathrm{HCN})/X(\mathrm{HNC})$) and their reported total integrated intensity ratios in our molecular maps (i.e., $I(\mathrm{HCN})/I(\mathrm{HNC})$) (see Sect.~\ref{sec:Ebarrier}).
This simple equality $X(\mathrm{HCN})/X(\mathrm{HNC})=I(\mathrm{HCN})/I(\mathrm{HNC})$, simplifies the otherwise dedicated conversion of each individual isomeric abundance estimate into its corresponding intensity value in our large molecular dataset ($>$~5000 positions).
While motivated by the similar excitation conditions of these two isomers \citep[e.g.][]{GOL86},
the direct correspondence between these abundance and intensity variations might be affected by distinct opacity and excitation effects.
In this appendix we explore the validity and limitations of this approximation for the comparison of both simulations and observations.

We simulated the expected total integrated intensity of the HCN and HNC molecules with a series of independent radiative transfer calculations using RADEX \citep{RADEX} adopting a characteristic linewidth of $\Delta V=$~1~km~s$^{-1}$ and a constant gas density of $n(\mathrm{H}_2)=10^4$~cm$^{-3}$ similar to our fiducial model 4 (see Sect.~\ref{sec:ratios}).
Assuming that the emission of both HCN and HNC isomers is generated from the same gas parcel, 
we explored different abundance ratios by comparing similar column density ratios of these two isomers, that is,  $N(\mathrm{HCN})/N(\mathrm{HNC})=X(\mathrm{HCN}))/X(\mathrm{HNC})$.
To reduce the number of free parameters, we obtained the expected intensity of HCN for different column density configurations and compared them with a prototypical HNC intensity in our cloud at different gas temperatures.
In particular, we simulated the total HNC integrated intensity for a fixed column density of $N(\mathrm{HNC})=10^{14}$~cm$^{-2}$, equivalent to a total gas column density of $N(\mathrm{H}_2)=10^{22}$~cm$^{-2}$ (or $A_\mathrm{V}\sim 10$~mag) for a typical absolute abundance of $X(\mathrm{HNC})\sim10^{-8}$ according to our chemical models (see Fig.~\ref{fig:AbsAbun}). At the same time, we obtained the total HCN intensity (including all hyperfine components) for distinct column densities between $N(\mathrm{HCN})=10^{13}$~cm$^{-2}$ and $10^{16}$~cm$^{-2}$.
Effectively, this comparison allowed us to explore three orders of magnitude in abundance differences with $N(\mathrm{HCN})/N(\mathrm{HNC})=X(\mathrm{HCN})/X(\mathrm{HNC})=[0.1,100]$.

Figure~\ref{fig:LvsX} (left) shows the results of our RADEX simulations 
for gas kinetic temperatures between $T_\mathrm{K}=$~10~K and 70~K (see coloured lines)
within a range of abundance variations between $X(\mathrm{HCN})/X(\mathrm{HNC})\sim$~1 and 20, in agreement to the observed intensity variations seen in Fig.~\ref{fig:Luminosities}.
Overall, our models show a positive correlation between their abundance and intensity ratios driven by the increasing HCN abundance. 
Nevertheless, the correspondence between these two quantities changes as a function of $X(\mathrm{HCN})/X(\mathrm{HNC})$. At low abundance ratios (e.g. $X(\mathrm{HCN})/X(\mathrm{HNC})\sim$~1), the observed $I(\mathrm{HCN})/I(\mathrm{HNC})$ intensity ratios follow an approximately 1:1 linear variation with respect to $X(\mathrm{HCN})/X(\mathrm{HNC})$, as expected in the optically thin regime (see dotted line). On the other hand, opacity and saturation effects reduce the equivalent HCN intensities and produce a much shallower correlation at increasing abundance ratios (i.e. $X(\mathrm{HCN})/X(\mathrm{HNC})>$~5). For a constant $T_\mathrm{K}$ value (e.g. $T_\mathrm{K}=10$~K, blue line), our RADEX models illustrate the increasing differences between the observed $I$(HCN)/$I$(HNC) intensity variations in comparison to their corresponding $X(\mathrm{HCN})/X(\mathrm{HNC})$ abundance ratios at high $X(\mathrm{HCN})/X(\mathrm{HNC})$ values. 

Interestingly, the reported connection between the gas kinetic temperature and the HCN production predicted by our chemical models (Sect.~\ref{sec:ratios}) improves the correspondence between $X(\mathrm{HCN})/X(\mathrm{HNC})$ and $L(\mathrm{HCN}))/L(\mathrm{HNC})$.
As Figure~\ref{fig:LvsX} (left) shows, for a given $X(\mathrm{HCN})/X(\mathrm{HNC})$ value   warmer gas temperatures
present higher $I$(HCN)/$I$(HNC) intensity values because the excitation and intensity of the HCN (1-0) line are enhanced. 
As a result, increasing gas temperatures at $X(\mathrm{HCN})/X(\mathrm{HNC})>$~2 can (partially) mitigate the saturation effects of opacity and produce $I$(HCN)/$I$(HNC) intensity values that are much closer to the linear dependence that is observed at lower $X(\mathrm{HCN})/X(\mathrm{HNC})$ values.
We illustrate these changes with the results of our model 4 (coloured dots in Fig.~\ref{fig:LvsX} left). While it still departs from a perfect linearity (see grey triangles), the expected correlation between the gas temperatures and abundance ratios (solid circles) reduces the differences between the intrinsic $X(\mathrm{HCN})/X(\mathrm{HNC})$ abundance variations and the observed $I$(HCN)/$I$(HNC) intensity variations down to $\lesssim$~0.3~dex throughout the range of $X(\mathrm{HCN}))/X(\mathrm{HNC})$ values we explored here (see dashed line).

Similar to our observational analysis (e.g. Fig.~\ref{fig:EBarrier}), in Figure~\ref{fig:LvsX} (right) we display the differences between the true abundance variations ($X(\mathrm{HCN})/X(\mathrm{HNC})$; solid lines) of model 4
with 1/$T_\mathrm{K}$ and the corrected  intensity ratios ($I$(HCN)/$I$(HNC); dashed lines) estimated by our RADEX calculations.
As illustrated by this comparison, the expected $I$(HCN)/$I$(HNC) values show a slightly shallower dependence with temperature than  $X(\mathrm{HCN})/X(\mathrm{HNC})$. As previously shown in Fig.~\ref{fig:LvsX}, these differences are more prominent at low 1/$T_\mathrm{K}$ values (i.e. high temperatures).
Nonetheless, these reported minor corrections are comparable to the local scatter of our observations ($\sim$~0.2~dex) and remain much smaller than the systematic temperature variations reported in our Orion observations ($>$~1~dex; grey crosses) (see Sect.~\ref{sec:Ebarrier}).
Although not identical, our results confirm the almost direct equivalence between the $X(\mathrm{HCN})/X(\mathrm{HNC})$ abundance and $I$(HCN)/$I$(HNC) intensity ratios within the physical conditions considered in this work.
For practical purposes, we therefore assume that these $X(\mathrm{HCN})/X(\mathrm{HNC})$ and $I(\mathrm{HCN})/I(\mathrm{HNC})$ measurements 
are equivalent quantities, at least in a first-order approximation.

\section{Chemical models. Properties, parameter space, and constraints}\label{sec:appendix1}

\begin{figure*}
	\centering
	\includegraphics[width=\linewidth]{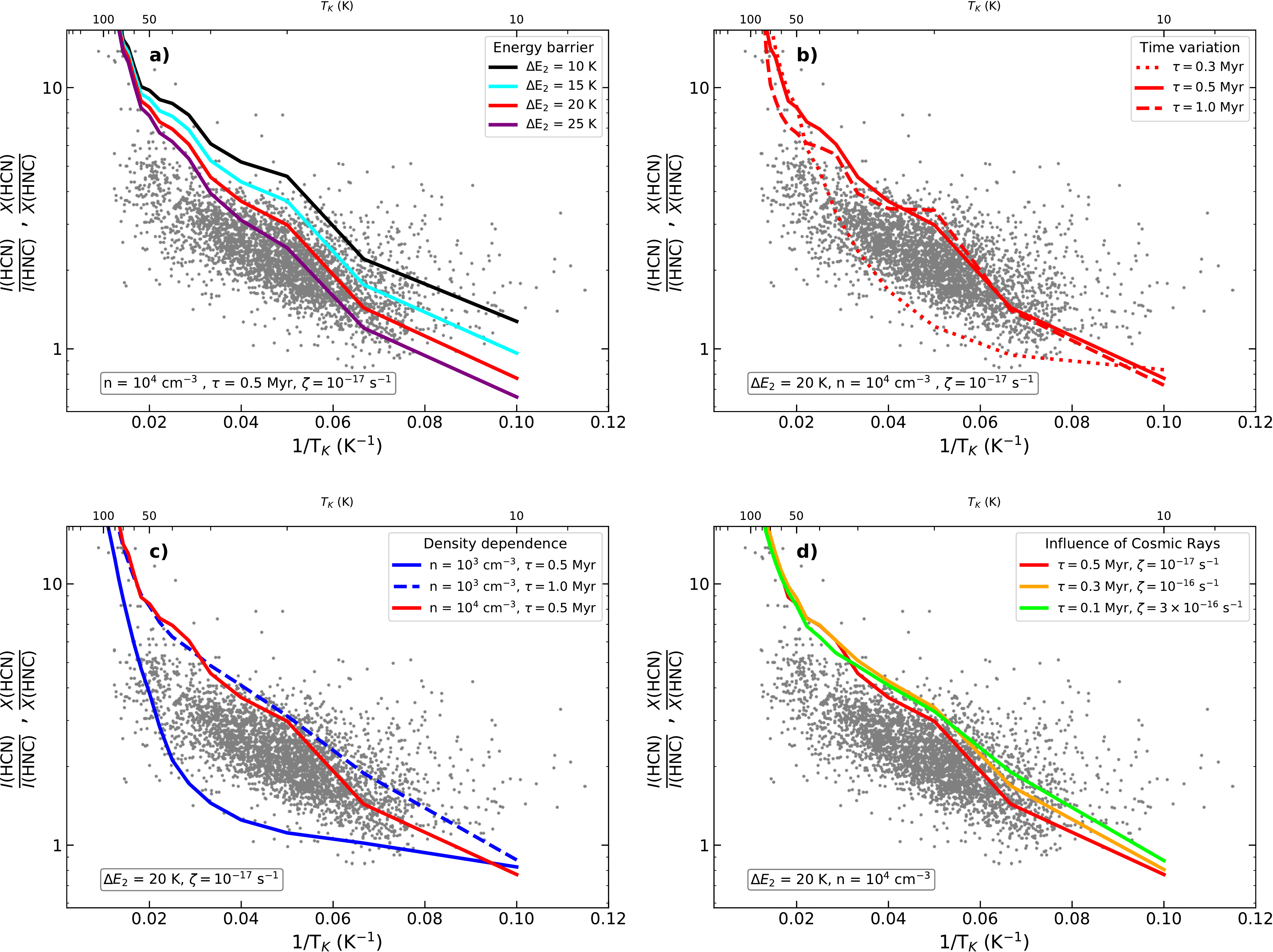}
	\caption{Predicted HCN and HNC abundances as a function of 1/$T_\mathrm{K}$ for different input parameters in our models.
		Data points and axes are similar to Fig.~\ref{fig:EBarrier}~b.
		From left to right and from top to bottom: model results for independent variations of the 
		{\bf (a)} energy barrier for reaction (2) ($\Delta E_2$),
		{\bf (b)} time evolution ($\tau$),
		{\bf (c)} gas density ($n$(H$_2$)), 
		and {\bf (d)} cosmic-ray ionization rate ($\zeta$).
		The input values of these models are listed and colour-coded in each plot in the upper right corner of each subplot. Similarly, all fixed parameters for each of these comparisons are indicated in the lower left corner.
		These results are compared to our fiducial model assuming $\Delta E_1=200$~K, $\Delta E_2=20$~K, $\tau = 0.5$~Myr, $n$(H$_2$)~=~10$^4$~cm$^{-3}$, and $\zeta=10^{-17}$~s$^{-1}$ (red solid line in all subplots).}
	\label{fig:models_evolution}
\end{figure*}

As demonstrated in Sect.~\ref{sec:Ebarrier}, our chemical models successfully reproduce the observed temperature dependence of the $I$(HCN)/$I$(HNC) intensity ratio assuming a fiducial model with energy barriers $\Delta E_1=200$~K and $\Delta E_2=20$~K for reactions (1) and (2), respectively, evolutionary timescales of $\tau = 0.5$~Myr, densities of $n$(H$_2$)~=~10$^4$~cm$^{-3}$, and cosmic-ray ionization rates of $\zeta=10^{-17}$~s$^{-1}$.
In this appendix we explore the potential effects of different input parameters in our chemical models in comparison to our fiducial case. When a constant $\Delta E_1=200$~K \citep{GRA14} is assumed, the four panels in Fig.~\ref{fig:models_evolution}(a-d) summarize different results of our simulations after varying each of the additional $\Delta E_2$, $\tau$, 
$n$(H$_2$), and $\zeta$ parameters independently. With these tests our goal is to investigate the properties and degeneracies of our models, as well as to quantify the errors in our parameter estimates.
 For simplicity, our comparisons assume a direct correspondence between the predicted variations on the abundance ratios in our models and the observed integrated intensities ratios, that is, $X(\mathrm{HCN})/X(\mathrm{HNC})=I(\mathrm{HCN})/I(\mathrm{HNC})$ (see Appendix~\ref{sec:appendix2}).

Figure~\ref{fig:models_evolution}a shows the effects of different energy barriers $\Delta E_2$ between 10~K and 25~K.  Changes in $\Delta E_2$ are directly translated into different slopes and offsets on the expected X(HCN)/X(HNC) abundance variations in our models within the effective temperature range of reaction (2), that is, at T$_K<$~40~K.
These variations are expected due to the exponential dependence of the observed X(HCN)/X(HNC) with $\Delta E_2$ (i.e. $\propto \mathrm{exp}\left( \frac{\Delta E_2}{\mathrm{T}_K} \right)$; Sect.~\ref{sec:Ebarrier}). In particular, lower $\Delta E_2$ values result in shallower temperature variations of the X(HCN)/X(HNC) abundance ratio. 
From the comparison with our observational data (grey points),
our fiducial model (i.e., $\Delta E_2$~=~20~K) presents a reasonable fit for both low and high temperatures simultaneously.
Nonetheless, variations of $\Delta(\Delta E_2)\pm$~5~K cannot be ruled out because of the intrinsic spread of our measurements. Additional analyses, including radiative transfer calculations and opacities, are needed to better constrain the precise value of this energy barrier $\Delta E_2$.

We investigate the time evolution of our chemical models in Figure~\ref{fig:models_evolution}b. We show the temperature dependence of the X(HCN)/X(HNC) ratio in three representative time stamps, namely, $\tau = 0.3$~Myr, $\tau = 0.5$~Myr, and $\tau = 1.0$~Myr. The results of this figure illustrate the characteristic evolutionary trend of our models. Early stages of evolution ($\tau = 0.3$~Myr; dotted line) are dominated by reaction (1) which produces a fast increase of the HCN abundance at high temperatures. This rapid evolution is followed by the comparatively slower effect of reaction (2)  ($\tau = 0.5$~Myr; solid line) which determines the slope of the X(HCN)/X(HNC) ratio at intermediate temperatures (see above). Later on ($\tau = 1.0$~Myr; dashed line), the combination of reactions (1) and (2) reach a steady state (or pseudo-equilibrium) in which the observed X(HCN)/X(HNC) dependence remains unaltered until the end of our simulations ($\tau = 3.0$~Myr, not shown). 

We explore the effects of the gas densities in our simulations in Figure~\ref{fig:models_evolution}c.  For simplicity, we show the results for the expected X(HCN)/X(HNC) values in two of our models with $n$(H$_2$)~=~10$^3$~cm$^{-3}$ (blue) evaluated at $\tau = 0.5$~Myr (blue solid line) and $\tau = 1.0$~Myr (blue dashed line).
As denoted by the comparison with our fiducial model (red solid line), lower densities result in a steeper temperature dependence of the X(HCN)/X(HNC) at similar timescales. Interestingly, the reported X(HCN)/X(HNC) variations for models with $n$(H$_2$)~=~10$^3$~cm$^{-3}$ and $\tau = 0.5$~Myr (blue solid line in Fig.~\ref{fig:models_evolution}c) resemble the functional dependence observed in models with  $n$(H$_2$)~=~10$^4$~cm$^{-3}$ and $\tau = 0.3$~Myr (red dotted line in Fig.~\ref{fig:models_evolution}b). Similarly, models with $n$(H$_2$)~=~10$^3$~cm$^{-3}$ and $\tau = 1.0$~Myr (blue dashed line) are similar to those with $n$(H$_2$)~=~10$^4$~cm$^{-3}$ and $\tau = 0.5$~Myr (fiducial model; red solid line). This time delay can be understood by the dependence of reaction (2) on the amount of free atomic O.  Controlled by the interaction between the number of molecules available (i.e. $n$(H$_2$)) and cosmic rays (see below), higher gas densities lead to higher absolute O densities in our models, which speeds up the effects of reaction (2) at low temperatures.

As for the density, the increase of the cosmic-ray ionization produces an enhancement of the atomic O abundances in our models, causing reaction (2) to act faster at low temperatures.
We illustrate the effects of cosmic rays in Figure~\ref{fig:models_evolution}d by displaying three different combinations of $\tau$ and $\zeta$ values. 
The selected models show indistinguishable results throughout the entire temperature range explored here. As a rule of thumb, increasing ionization rates $\zeta$ allow our chemical models to reach their steady state (fiducial model) in shorter timescales $\tau$, satisfying an approximate relationship of $(\tau \times \zeta) \sim$~const.

Each of these models considers the chemical evolution of an idealized portion of the ISM that is described by a single combination of $\tau$, $n$(H$_2$), and $\zeta$ parameters. 
Our individual tests reveals some of the intrinsic degeneracies of our models (e.g.  $n$(H$_2$) \& $\zeta$ vs. $\tau$).
These oversimplified simulations contrast with the definitely more complex physical conditions found in real clouds. In particular, observations in regions such as Orion undoubtedly sample a wide range of gas regions including a large variety of densities, evolutionary timescales, and radiation fields (e.g. see maps in Fig.~\ref{fig:ISF_maps}). 
The combination of these effects is likely responsible for the spread reported in our $I$(HCN)/$I$(HNC) measurements. 
Despite all these caveats, the predicted HCN-to-HNC ratios in our models show consistent results within $\lesssim$~0.4~dex for variations of at least one order of magnitude in all $\tau$, $n$(H$_2$), and $\zeta$ parameters. 
In this context, our fiducial model is meant to represent the average temperature dependence of the X(HCN)/X(HNC) abundance ratio under average ISM conditions.
Nonetheless, additional simulations, including realistic density, chemistry, and irradiation conditions as well as detailed radiative transfer calculations \citep[e.g.][]{PEN18,SEI19}, are needed to fully quantify the observed X(HCN)/X(HNC) abundance variations in higher detail.

Our parameter space exploration highlights the dominant role of reaction (2) at temperatures of T$_K\lesssim$~40~K. After a short activation period, this reaction (2) reaches a steady state in all our models solely determined by its energy barrier $\Delta E_2$ (Fig.~\ref{fig:models_evolution}a).
The activation timescales for reaction (2) (typically of $\tau \sim $~0.5~Myr) are determined by the amount of atomic O available in the gas phase generated by cosmic-ray ionization (see Fig.~\ref{fig:models_evolution}c and d). 
After it is established, however, this stable configuration remains unaltered over timescales of several Myr (Fig.~\ref{fig:models_evolution}b).
This pseudo-equilibrium, independent of the initial conditions in our models, would explain the strong correlation between $T_\mathrm{K}$ and $I$(HCN)/$I$(HNC) that is observed in a wide range of ISM environments in our Galaxy (see also Sect.~\ref{sec:universal}).

\end{appendix}

\end{document}